\documentclass[12pt]{article}
\usepackage{amsfonts, graphicx}
\usepackage{amsmath}
\usepackage{multirow}
 \usepackage[authoryear,comma]{natbib}
\usepackage{autonum}

\usepackage{amstext} 
\usepackage{array}   
\newcolumntype{R}{>{$}r<{$}} 

\newtheorem{theorem}{Theorem}
\usepackage{xr}

\providecommand{\keywords}[1]{\textbf{\textit{Keywords:}} #1}

\title{The B-Exponential Divergence and its Generalizations with Applications to Parametric Estimation}

\author{Taranga Mukherjee$^1$, Abhijit Mandal$^2$, Ayanendranath Basu$^3$\\ \\
 $^1$University of Calcutta, Kolkata, W.B., India\\
$^2$Wayne State University, Detroit, MI, USA\\
$^3$Indian Statistical Institute, Kolkata, W.B., India}

\date{}

\begin{document}

%
%
%

\maketitle

\begin{abstract}
In this paper a new family of minimum divergence estimators based on the Br{\`e}gman  divergence is proposed, where the defining convex function has an exponential nature. These estimators avoid the necessity of using an intermediate  kernel density and many of them  also have strong robustness properties. It is further demonstrated that the proposed approach can be extended to construct a class of generalized estimating equations, where the pool of the resultant estimators encompass a large variety of minimum divergence estimators and  range from highly robust to fully efficient based on the choice of the tuning parameters. All of the resultant estimators are M-estimators, where the defining functions make explicit use of the form of the parametric model. The properties of these estimators are discussed in detail; the theoretical results are substantiated by simulation and real data examples. It is observed that in many cases, certain robust estimators from the above generalized class provide better compromises between robustness and efficiency compared to the existing standards. 
\end{abstract}






\section{Introduction}
In parametric estimation there are two fundamental but potentially competing ideas -- efficiency when the model is appropriately chosen and robustness when it is not. Mathematically, the trade-off between efficiency and robustness is inevitable if
one adheres to the notion of robustness based on the influence function analysis. Under standard regularity conditions,  the maximum likelihood estimator (MLE) has the highest asymptotic efficiency  among all classes of estimators. However, the MLE is, almost always, an extremely poor performer in terms of robustness.  \cite{MR0448700} introduced an estimation procedure using the Hellinger distance, which generates the minimum Hellinger distance estimator (MHDE). This estimator has significantly better robustness properties compared to the MLE, but shares the property of full asymptotic efficiency at the model. In spite of such  nice properties, this procedure  has a major drawback in that it requires the use of a nonparametric kernel density estimator.  \cite{MR1665873} introduced a  robust and efficient minimum divergence estimation procedure for parameter estimation which does not require any such nonparametric smoothing. This process is based on the density power divergence (DPD), a family of density based divergence measures, where the divergences represent  special cases of the Br{\`e}gman divergence (\citealp{Bregman1967200}). In the present paper we have also followed the Br{\`e}gman strategy to develop a new class of estimators,  indexed by a real tuning parameter $\alpha$.

There are many different estimation methods in statistics, and even within the field of robust density-based minimum distance estimators, several good choices are available. In presenting a new set of estimators, therefore, it is imperative to show that new estimators are competitive with, or preferably better, than the existing standard. In the class of minimum distance estimators which bypass the need for nonparametric smoothing, the minimum DPD estimators represent the current standard. In this paper we will show that the generalized estimating equations originating from the Br{\`e}gman divergence represent a class of estimators, which includes the minimum DPD estimators as a special case, but often provide us with many better options. 


 In Section \ref{Bregman} we  introduce the new divergences, called B-exponential divergence, in the  spirit of the Br{\`e}gman divergence; the corresponding minimum divergence estimation procedure is  also described in this section. A class of generalized estimating equations encompassing the  minimum B-exponential divergence estimation and the  minimum DPD estimation is proposed in Section \ref{sec:gen}. In Section \ref{sec:properties} we discuss the theoretical properties  of this new class of estimators.  Simulation results and real data examples are given in Sections \ref{sec:simulation} and \ref{sec:real_data}, respectively. In Section \ref{sec:tuning} the optimal values of the tuning parameters are explored.  The estimation method is extended to the case of the regression model in Section \ref{sec:reg}. Finally,  in Section \ref{sec:conclusion} some concluding remarks are presented. Additional numerical results and a sketch of the proof are given in the Supplementary Material.  
 
\section{The B-Exponential Divergence} \label{Bregman}
The Br{\`e}gman divergence (\citealp{Bregman1967200}), used in convex programming,  was  introduced  to define a distance between two points in $\mathbb{R}^d$. The general form of the Br{\`e}gman divergence between two densities $g$ and $f$ is
\begin{equation}
D_B(g,f)=\int_x \left[ B(g(x))-B(f(x))-(g(x)-f(x))B^\prime(f(x))\right] dx,
\label{breg}
\end{equation}
 where  $B: \mathbb{R} \rightarrow \mathbb{R}$ is a convex function, and $B'$ is the derivative of $B$.
 Note that the function $B(y)$ is not uniquely determined as $B(y)$ and $B^*(y)=B(y)+ay+b$ generate the same divergence for any real numbers $a$ and $b$. 
 
 The density power divergence, introduced by \cite{MR1665873}, is given by
\begin{equation}
P_\beta(g,f)=\int_x \left[ f^{\beta+1}(x)-\left(1+\frac{1}{\beta}\right)g(x)f^\beta(x)+\frac{1}{\beta}g^{\beta+1}(x)\right] dx, \ \beta >0 .
\end{equation}
 It is a special case of the Br{\`e}gman divergence when the defining convex function has the form $B(y)=\frac{y^{\beta+1} - y}{\beta}$ (see \citealp{MR1859416}). The  density power divergence for $\beta=0$  is defined by its limiting case as $\beta \rightarrow 0$, which turns out to be the Kullback-Leibler divergence
  \begin{equation}
 P_0(g,f)=\int_x g(x)\log\left( \frac{g(x)}{f(x)}\right)dx. \label{kl}
\end{equation}
In this case   $B(y) = y\log y$. Another special case corresponds to $\beta=1$ or $B(y)=y^2$, and in this case  the divergence is the (squared) $L^2$ distance 
\begin{equation}
P_1(g,f)=
\int_x (g(x) -f(x))^2 dx. 
\label{l2_dist}
\end{equation}
%

We now propose our new divergence called the Br{\`e}gman-exponential divergence or, in short, the B-exponential divergence (BED), by taking 
\begin{equation}
B(y)=\frac{2 (e^{\alpha y}-\alpha y -1)}{\alpha^2}. 
\end{equation}
 The form of the divergence is given by
\begin{equation}
d_{\alpha}(g,f)= \frac{2}{\alpha}
\int_x \left[e^{\alpha f(x)}\left(f(x) -\frac{1}{\alpha}\right)  - 
e^{\alpha f(x)} g(x) + \frac{1}{\alpha }  e^{\alpha g(x)} \right] dx, \ \ \alpha \in {\mathbb R}. \label{expdpd}
\end{equation}
 The divergence for $\alpha=0$ is defined by its limiting case as $\alpha \rightarrow 0$, which turns out to be the (squared) $L^2$ distance. In fact, the multiplier $\frac{2}{\alpha}$ is attached to the distance to ensure that the squared $L^2$ distance is a part of this family.  However, Kullback-Leibler divergence does not belong to this family for any value of $\alpha$. 


\subsection{The Estimating Equation}\label{SEC:expdpd}
Consider a random variable $X$ with unknown distribution function $G$.  Let $g$ represent the corresponding probability density function.  We choose a parametric family of densities
$\{f_\theta:\theta\in\Theta\subset {\mathbb{R}}^{p}\}$ to model the true density $g$, and  $\theta$ is the parameter of interest.  The minimum B-exponential  divergence
functional $T_\alpha(G)$ at $G$ is defined by the relation $d_{\alpha
}(g,f_{T_\alpha(G)})=\min_{\theta\in\Theta}d_{\alpha
}(g,f_\theta)$. The term $\int_x e^{\alpha g(x)} dx$ has no role in the
minimization of $d_\alpha(g,f_\theta)$ over
$\theta\in\Theta$, and may be ignored in the minimization process. Thus, the essential objective function
to be minimized in the computation of  $T_\alpha(G)$ reduces, for $\alpha \neq 0$, to 
\begin{equation}
\int_xe^{\alpha f_\theta(x)} \left(f_\theta(x) -\frac{1}{\alpha}\right)dx -  \int_x e^{\alpha f_\theta(x)} g(x)dx .
\label{bed}
\end{equation}
Given a random
sample $X_{1},\ldots,X_{n}$ from  $G$, we can approximate the second term of the 
 objective function in (\ref{bed}) by the empirical mean of $ e^{\alpha f_\theta(x)}$.
For a given tuning parameter $\alpha (\neq 0)$, therefore, the minimum B-exponential
divergence estimator $\widehat\theta_\alpha = T_\alpha(G_n)$ of
$\theta$ can be obtained by minimizing
\begin{equation}
\frac{1}{n} \sum_{i=1}^n V_\theta(X_i) =
 \int_xe^{\alpha f_\theta(x)}\left(f_\theta(x) -\frac{1}{\alpha}\right) dx - 
\frac{1}{n} \sum_{i=1}^n e^{\alpha f_\theta(X_i)} 
\label{EQ:Vtheta}
\end{equation}
over $\theta\in\Theta$, where $V_{\theta%
}(x)=\int e^{\alpha f_\theta(y)}\left(f_\theta(y) -\frac{1}{\alpha}\right) dy-e^{\alpha f_\theta(x)}$. The minimization avoids the use of
a nonparametric density estimate. 

Let $u_\theta(x) = \frac{\partial}{\partial \theta}
\log f_\theta(x)$ be the likelihood score function of the model.
Under differentiability of the model with respect to $\theta$, the optimization of the objective
function in Equation (\ref{EQ:Vtheta})
leads to an estimating equation of the form
\begin{equation}
\frac{1}{n}\sum_{i=1}^{n}u_\theta(X_{i}) f_\theta(X_{i}) e^{\alpha f_\theta(X_{i})}
-\int_x u_\theta(x)f_\theta^2(x)e^{\alpha f_\theta(x)}  dx=0, \label{EQ:general_case}%
\end{equation}
which is an unbiased estimating equation under the model. 
The minimum B-exponential divergence estimator (MBEDE) is obtained as a solution of the estimating equation in (\ref{EQ:general_case}).

The above estimating equation can be viewed in the spirit of a  weighted likelihood estimating equation. Since the  estimating equation weights the score $u_\theta(X_i)$
with $f_\theta(X_{i}) e^{\alpha f_\theta(X_{i})}$, the outlier resistant behavior of the estimator is
intuitively apparent. The weights are plotted in Figure \ref{fig:weights} in the Supplementary Material for different values of $\alpha$ when $f_\theta$ is  the standard normal distribution.
For $\alpha=0$, the estimating equation weights the score with  density $f_\theta$. When $\alpha \neq 0$, the weight function downweights the tail values more sharply when $\alpha>0$, while the weights are  relatively more uniform when $\alpha<0$. 
See \cite{MR1665873} and \cite{MR1859416} for other examples of  density based  minimum divergence estimators in the same spirit.

\section{A Generalized Estimator} \label{sec:gen}
Consider the estimating equation of the MBEDE given by Equation (\ref{EQ:general_case}). 
Unlike the DPD, the family of B-exponential divergences does not include the Kullback-Leibler divergence as a special case. So, none of the MBEDEs have full asymptotic efficiency at the model. In the normal mean problem, for example,  the most efficient MBEDE corresponds to approximately $\alpha =- 7.44$ with an asymptotic relative efficiency (ARE) of 88.48\% (see Section \ref{sec:simulation} for details). However, the estimating equation in (\ref{EQ:general_case}) can easily be extended to construct a more general estimating equation having the form  
\begin{equation}
\frac{1}{n}\sum_{i=1}^{n}u_\theta(X_{i}) f_\theta^\beta(X_{i}) e^{\alpha f_\theta(X_{i})}
-\int_x u_\theta(x)f_\theta^{1+\beta}(x)e^{\alpha f_\theta(x)}  dx=0, \label{EQ:new}%
\end{equation}
where $\alpha \in \mathbb{R},\ \beta \geq 0$.
Although the estimating equation (\ref{EQ:new}) does not exactly correspond to a minimum divergence method (the corresponding divergence does not exist in closed form), we will denote the associated estimator as the generalized B-exponential divergence estimator, and refer to it as GBEDE($\alpha, \beta$). In the special case when $\alpha=0$, the minimum density power divergence estimator (MDPDE) with tuning parameter $\beta$  represents a solution of the GBEDE($0, \beta$) estimating equation, while the MBEDE with parameter $\alpha$ is a solution of the GBEDE($\alpha, 1$) estimating equation.
For $\alpha=\beta=0$, Equation (\ref{EQ:new}) reduces to the ordinary likelihood score equation, having the MLE as a solution. This two parameter family represented by Equation (\ref{EQ:new}) gives us  more flexibility to come up with an estimator having a better balance between robustness and efficiency.
As this  generalized super class of estimators GBEDE contains both the  MDPDEs and the MBEDEs classes,  a study of the estimators within this generalized class automatically entails a wide comparison involving these two families and much more.

Although there does not exist any explicit divergence corresponding to the estimating equation (\ref{EQ:new}), one may find an approximate measure  based on the numerical integration. For $\alpha<0$ and $x>0$, it can be shown that 
\begin{equation}
\int_x x^\beta e^{\alpha x} dx = - \frac{  \Gamma(\beta+1, -\alpha x)}{(-\alpha) ^{\beta+1}},
\label{int}
\end{equation}
where $\Gamma(a, y) = \int_y^\infty t^{a-1} e^{-t} dt$ is the  `upper'  incomplete gamma function. Let us denote the  function in Equation (\ref{int}) as $\xi(x, \alpha, \beta)$. Then, up to a constant term, the empirical  divergence corresponding to the estimating equation (\ref{EQ:new}) is given by
\begin{equation}
\int_x \xi(f_\theta(x), \alpha, \beta) dx - \frac{1}{n}\sum_{i=1}^n \xi(f_\theta(X_i), \alpha, \beta-1). \label{div_gen}
\end{equation}
In case the estimating equation in (\ref{EQ:new}) exhibits multiple roots, one can select the estimator to be the root that minimizes the above function. An example is provided in the Supplementary Material. 

\section{Properties} \label{sec:properties}
In this section we study the important theoretical properties of the GBEDEs. 

\subsection{Link with M-estimation}
The generalized  B-exponential  divergence estimator is an M-estimator (see \citealp{MR606374}, \citealp{MR829458}). For an M-estimator, the estimating equation is written as $\sum_i \psi(X_i, \theta) = 0$, for a suitable function $\psi(\cdot,\cdot)$ from $\mathbb{R}^2$ to $\mathbb{R}$. From Equation (\ref{EQ:new}) we obtain the $\psi$ function for the GBEDE as
\begin{equation}
 \psi(x, \theta) = u_\theta(x) f^\beta_\theta(x) e^{\alpha f_\theta (x)} - \int_y u_\theta(y)f_\theta^{1+\beta}(y)e^{\alpha f_\theta(y)}  dy .
 \label{phi_fun}
\end{equation}
Notice that this is an unbiased estimating equation under the model. Therefore, the theoretical properties of the GBEDE can be derived directly from the properties of M-estimators.

\subsection{The Asymptotic Distribution\label{SEC:asymp_dist1}}

Let $g$ be the true data generating density. By exploring the theoretical estimating equation 
$\int \psi(x, \theta)dG(x) = 0$ where $\psi(\cdot, \cdot)$ is as in  (\ref{phi_fun}), it is immediately seen that the functional $T_{\alpha,\beta}(G)$ associated with the GBEDE($\alpha, \beta$)   is Fisher consistent; it recovers the true value of $\theta$ when
the data generating density  is in the model, i.e. $g=f_\theta$ for some $\theta \in \Theta$. When $g$ is not in the model,   $\theta_{\alpha,\beta}=T_{\alpha,\beta}(G)$ will be the root of the theoretical version of the generalized estimating equation (\ref{EQ:new}) which may be expressed as 
\begin{equation}
\int u_\theta(x) f_\theta^\beta(x) e^{\alpha f_\theta(x)} g(x) dx - 
\int  u_\theta(x) f_\theta^{1+\beta}(x) e^{\alpha f_\theta(x)} dx = 0. 
\end{equation} The density function $f_{\theta_{\alpha,\beta}}$  represents the model element
closest to $g$ in the GBEDE sense.  Let us define
\begin{equation}
\begin{split}
 &J_{\alpha,\beta}(\theta) = \int_x u_\theta(x) u_\theta^T(x) f_\theta^{1+\beta}(x) e^{\alpha f_\theta(x)} dx\\
&+\int_x\Big\{I_\theta(x)- \beta u_\theta(x)u_\theta^T(x) - \alpha u_\theta(x)u_\theta^T(x) f_\theta(x)\Big\}
\Big\{g(x)-f_\theta(x)\Big\}f_\theta^\beta(x) e^{\alpha f_\theta(x)} dx 
\label{EQ:J-matrix1}
\end{split}
\end{equation}
and
\begin{equation}
K_{\alpha,\beta}(\theta)=\int_x u_\theta(x)u_\theta^T(x)f_\theta^{2\beta}(x) e^{2 \alpha f_\theta(x)} g(x)dx-\xi_{\alpha,\beta}(\theta)\xi_{\alpha,\beta}^T%
(\theta) , \label{EQ:K-matrix1}%
\end{equation}
where
\begin{equation}
 \xi_{\alpha,\beta}(\theta)=\int_x u_{\theta%
}(x)f_\theta^\beta(x) e^{\alpha f_\theta(x)} g(x)dx, 
\label{xi1}
\end{equation}
and  $I_{\theta}(x)=-\frac{\partial}{\partial\theta}u_\theta(x)$,
the  information function of the model.
%
%
The following theorem provides the asymptotic distribution of the GBEDE.

\begin{theorem} \label{theorem1}
Assume that the conditions D1--D5 in the Supplementary Material are satisfied. These conditions are the obvious generalizations of the conditions of \cite{MR2830561}, p. 304. Then
   
\begin{enumerate}
\item[a)] the  estimating equation 
(\ref{EQ:new}) has a consistent sequence of roots
$\widehat\theta_{\alpha,\beta}=T_{\alpha,\beta}(G_n)$,

\item[b)] $n^{1/2}(\widehat\theta_{\alpha,\beta}-\theta_{\alpha,\beta})$ has an asymptotic multivariate normal distribution with mean
zero (vector) and covariance matrix $J_{\alpha,\beta}^{-1}K_{\alpha,\beta}J_{\alpha,\beta}%
^{-1}$, 
where $J_{\alpha,\beta}$ and $K_{\alpha,\beta}$ are as in Equations
(\ref{EQ:J-matrix1}) and (\ref{EQ:K-matrix1}) respectively.
\end{enumerate}
\label{theorem:asymp}
\end{theorem}
When the true distribution $G$ belongs to the model so that
$G = F_\theta$ for some $\theta \in \Theta$, the formula for
$J = J_{\alpha,\beta}(\theta), K =
K_{\alpha,\beta}(\theta)$ and $\xi = \xi_{\alpha,\beta}(\theta)$
simplify to
\begin{equation}
\begin{split}
& J = \int_x u_\theta(x)
u_\theta^T(x) f_\theta^{1+\beta}(x) e^{\alpha f_\theta(x)} dx,
~~K = \int_x u_\theta(x)
u_\theta^T(x) f_\theta^{1+2\beta} (x) e^{2 \alpha f_\theta(x)}  dx
- \xi \xi^T, \\
& \xi = \int_x u_\theta(x) f_\theta^{1+\beta} (x) e^{\alpha f_\theta(x)}  dx. \label{model_variance1}
\end{split}
\end{equation}

\begin{table}
\caption{Asymptotic relative efficiency of GBEDE under the normal location model.}
\begin{center}
\setlength{\tabcolsep}{4pt}
\begin{tabular}{r|rrrrrrrrrrr}
\hline
$\alpha \backslash \beta$ &0&0.1  & 0.2 &0.3 & 0.4&0.5  & 0.6&0.7&0.8 & 0.9 & 1\\ 
\hline
0 & 100.00 &  98.76 &  95.86 &  92.11 &  88.00 &  83.80 &  79.66 & 75.67 &  71.88 &  68.30 &  64.95 \\

$-1$ & 98.99 &    99.62 & 98.19 &    95.55 & 92.24 & 88.57 & 84.78 & 80.99 &  77.28 & 73.71 & 70.29 \\

$-2$ &  92.46 & 96.21 & 97.65 & 97.50 & 96.24 & 94.22 & 91.71 & 88.88 &  85.86 & 82.77 & 79.65 \\

$-3$ &     87.98 & 92.77 & 95.36 & 96.32 & 96.09 & 94.99 & 93.28 & 91.12 &  88.66 & 86.00 & 83.24 \\
\hline
\end{tabular}\label{normal_mu_are}
\end{center} 
\end{table}

Note that the matrix $J$ is the expectation of the partial derivative matrix, and $K$ is the covariance matrix, of the $\psi$ function given in Equation (\ref{EQ:J-matrix1}). We use the above formulas to evaluate the theoretical asymptotic efficiencies for the GBEDEs for the normal location parameter. The results are presented in Table \ref{normal_mu_are}. At $\alpha = 0$, the estimators are simply the minimum DPD estimators, and hence the asymptotic efficiencies are decreasing with  $\beta$. The situation changes for larger (in negative magnitude) values of $\alpha$. In such cases, for every fixed value of $\alpha$, the efficiencies initially rise with $\beta$ before reaching a peak and then drop  again. 

On the other hand, a review of these numbers for fixed values of $\beta$ reveals that the efficiencies of the other GBEDEs are greater than those of the MDPDEs for all $\beta \geq 0.3$. For larger values of $\beta$, this gain is substantial. This is a significant observation. Within the MDPDE class, larger values of $\beta$ are known to provide stability against data contamination, but at the cost of some loss in efficiency. This loss can assume very high proportions (compare the efficiency of 100\% at $\beta = 0$ to the efficiency of 64.95\% at $\beta = 1$) for large values of $\beta$. Much of this lost efficiency is recovered when one couples a large negative value of $\alpha$ with the same value of $\beta$. From the simulation results presented later it appears that the increase (in negative magnitude) in the tuning parameter $\alpha$ does not  affect the robustness properties in any significant manner. Thus the GBEDE family gives us several options for improving the efficiency of the robust MDPDEs, without compromising their stability.

\subsection{Influence function}
\begin{figure}
\centering
\caption{Influence functions of different GBEDEs in the $N(\mu,1)$ model.}
\includegraphics[height=8cm,width=12cm]{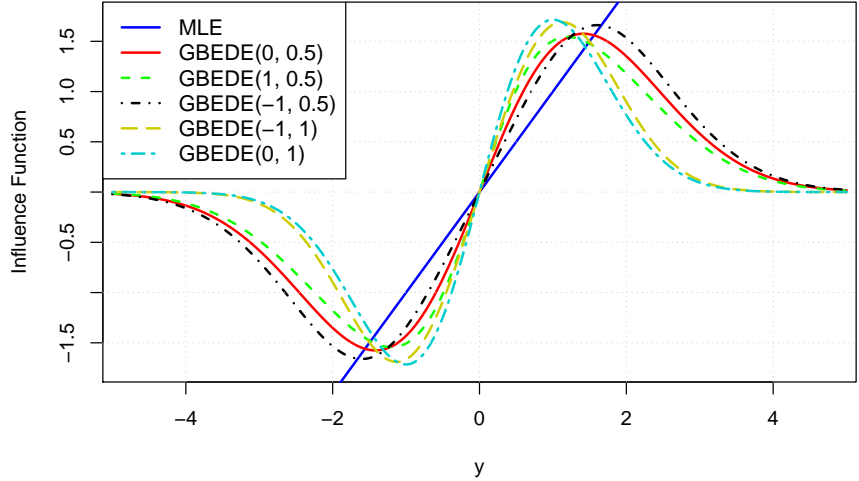}
\label{fig:inf}
\end{figure}

As GBEDEs are M-estimators, the influence function of these estimators is readily available. In the general setup this is given by 
\begin{equation}
 IF(y) = J_{\alpha,\beta}^{-1}(\theta) \left\{ u_\theta(y) f_\theta^\beta(y)  e^{\alpha f_\theta(y)}  - \xi_{\alpha,\beta}(\theta) \right\},
\end{equation}
where $J_{\alpha,\beta}$ and $\xi_{\alpha,\beta}$ are given in Equations (\ref{EQ:J-matrix1}) and (\ref{xi1}), respectively.  When the true distribution belongs to the model, 
a simplified form for the influence function is obtained by replacing $J_{\alpha,\beta}$ with $J$ and $\xi$ as given 
 in Equation (\ref{model_variance1}). 
If we assume that $J$ and $\xi$ are finite, then the influence function is bounded whenever $u_\theta(y) f^\beta_\theta(y)  e^{\alpha f_\theta(y)} $ is bounded in $y$. This is true for the most standard parametric models, including the normal location scale model, for all finite values of $\alpha$ and $\beta> 0$.  On the other hand, the influence function is not bounded in case of the maximum likelihood estimator (corresponding to $\alpha = \beta = 0$). In Figure \ref{fig:inf}, we have plotted the influence functions of GBEDEs for the mean parameter in the normal model when the scale parameter $\sigma$ equals 1. The  redescending nature of the influence functions of the GBEDE is clearly observed.  Here, ${\rm GBEDE}(0, 0.5)$ is the MDPDE with $\beta=0.5$, ${\rm GBEDE}(-1, 1)$ is the MBEDE with $\alpha=-1$, and ${\rm GBEDE}(0, 1)$ is the minimum $L_2$ distance estimator  or the MBEDE with $\alpha=0$.  

\section{Simulation Results} \label{sec:simulation}
\begin{figure}
\begin{center}
\begin{tabular}{rr}
 \includegraphics[height=5cm, width=7.5cm]{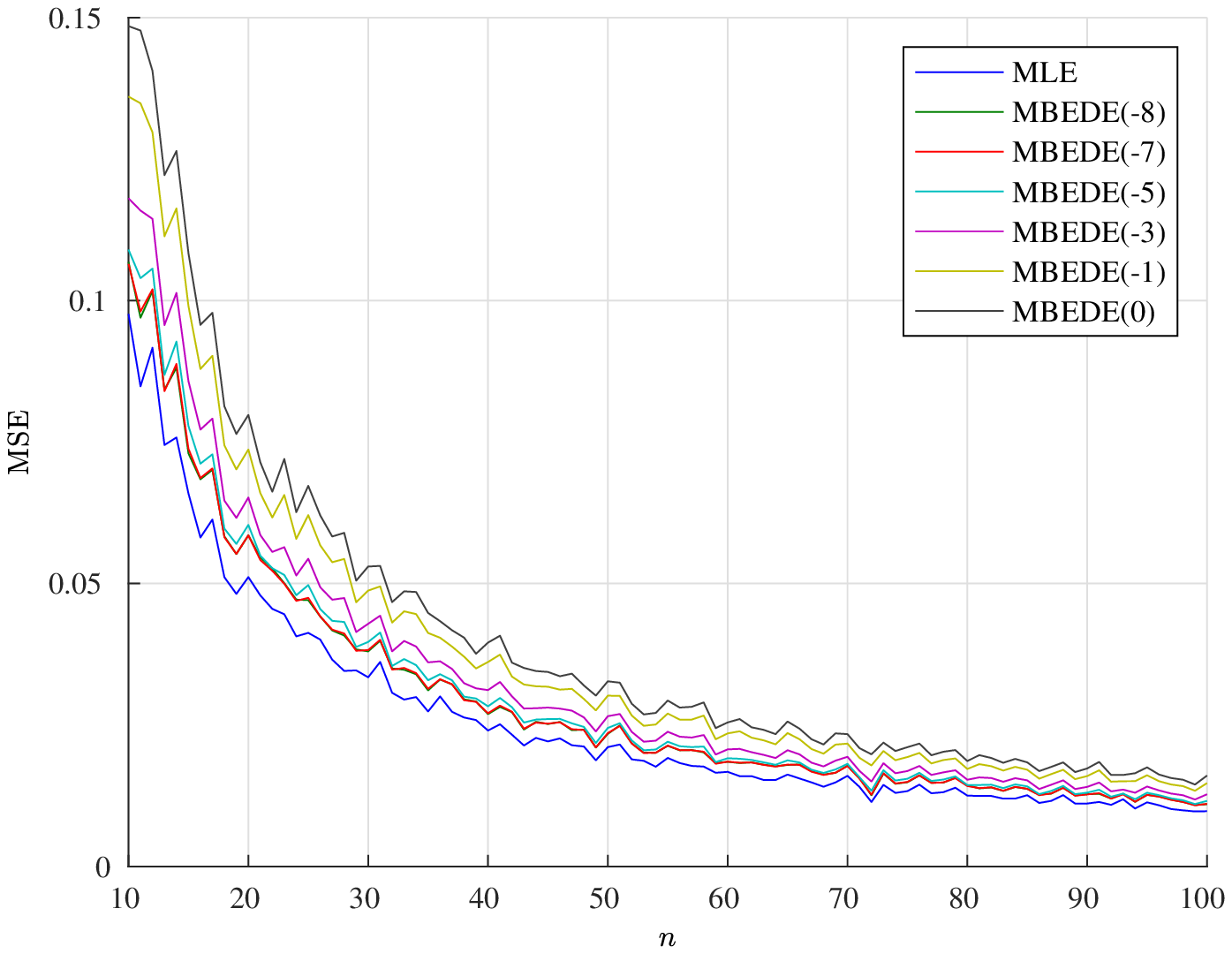}
 \negthinspace &
\negthinspace \includegraphics[height=5cm, width=7.5cm]{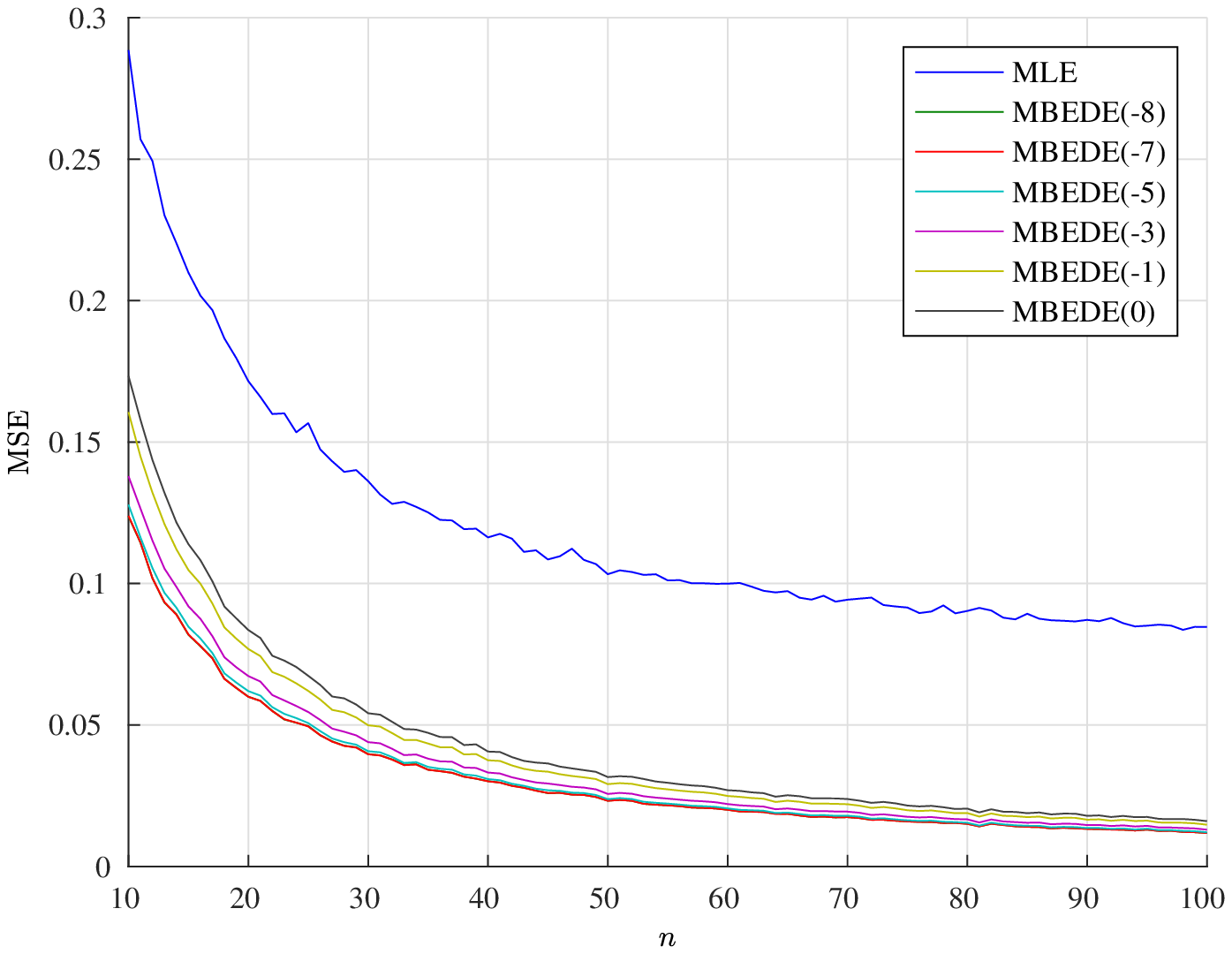}
 \\
\multicolumn{1}{c}{\textbf{(a)}} & \multicolumn{1}{c}{\textbf{(b)}} \\
\end{tabular}
\caption{MSEs  of different MBEDEs  in (a) pure  and (b) contaminated normal data. }
\label{fig:normal3}
\end{center}
\end{figure}

We begin with  some numerical studies to describe the performance of the  minimum  BED estimators.
 In the first experiment data are randomly generated from the standard normal distribution. We have assumed the $N(\mu, 1)$ model, and estimated the  mean parameter. The
MLE and MBEDEs of $\mu$ with $\alpha=0,-1, -3, -5,-7$ and $-8$ are considered. The MBEDEs are computed by minimizing the  B-exponential objective function in Equation  (\ref{EQ:Vtheta}) using the `fminsearch' function in Matlab. The empirical mean square error (MSE) of each estimator is calculated using 10,000 replications at each sample size between  $n=10$ to $n = 100$.  
The MSEs (multiplied by $n$)
are plotted against the sample size, and presented in the same frame   in Figure \ref{fig:normal3}(a).  
It is observed that the MSEs of the estimators stabilize around their corresponding asymptotic variances pretty early. It is also seen that  the MSEs of all the MBEDEs  considered are uniformly higher compared to that of the MLE for any value of  $\alpha$. The MSE decreases sharply as $\alpha$ increases (in negative magnitude), and it reaches minimum approximately at $\alpha =- 7.44$ with an asymptotic relative efficiency (ARE) of 88.48\%. (The curves corresponding to $\alpha = -7$ and $\alpha = -8$ are practically superimposed on each other). Then again it increases, however, the rate of increment is very slow in this case.

To evaluate the robustness properties of the MBEDEs under
contamination, we repeated the previous experiment with data generated from the normal mixture of 
$0.95N(0,1)+0.05 N(5,1)$. The data are still modeled by a parametric $N(\mu, 1)$ distribution.
 The MSEs, computed against the target value of 0 (the mean of the larger component of the mixture),  are now given in Figure \ref{fig:normal3}(b). For the values of $\alpha$ considered here,  the MBEDEs strongly downweight the effect of outliers. It is also interesting to note that the set of $\alpha$ values which minimize the mean square error under contamination ($\alpha = -5$ and $\alpha = -7$) are among the more efficient MBEDEs under pure data, at least as far as this example is concerned. 
On the whole, it appears that the efficiency of many of our MBEDEs are reasonably close to
the efficiency of the MLE in this model, but the robustness properties of the former are significantly better. Yet, since there is no member of this family which attains full asymptotic efficiency, it indicates that there is room for further refining this family to obtain maximum likelihood as a limiting case. The class of GBEDEs is developed with this intent. 

We now expand the scope of our  simulation to explore the efficiency of the GBEDEs, as obtained  by solving Equation  (\ref{EQ:new}).  We consider the two parameter normal model $N(\mu, \sigma^2)$, and the Poisson model with parameter $\lambda$, denoted as $P(\lambda)$. For the normal model and the Poisson model, the true data generating distributions are $N(0, 1)$ and $P(2)$,  respectively. In each case we have generated random samples of size $n = 100$, calculated the GBEDEs for different combinations of ($\alpha, \beta$) values, and replicated the process 10,000 times.  The GBEDEs are computed by solving the generalized estimating equation given in  (\ref{EQ:new}) using the `fsolve' function in Matlab. Tables \ref{normal_mu_pure}, \ref{normal_sig_pure} and \ref{poi_pure} in the Supplementary Material present the empirical relative efficiency of the estimators with respect to the MLE, where the MSEs (times $n$) of the MLEs of $\mu$, $\sigma$ and $\lambda$  are 1.01, 0.49 and 1.97, respectively. For pure data, the MLE, naturally, has maximum efficiency (minimum mean square error). However, several of the GBEDEs have relative efficiency very close to 1.   The row corresponding to $\alpha=0$ in each table represents the efficiency of the minimum density power divergence estimators (MDPDEs) for different values of $\beta$. For a fixed $\beta$, the GBEDEs which show equal or better relative efficiency than the corresponding MDPDE are marked in bold fonts  in these tables. It is noticed that in most of the cases the efficiency of the GBEDE may be improved, often significantly,  over the MDPDE if we set $\alpha$ at some appropriate negative value. 
This is an extremely significant observation, as this demonstrates that the introduction of this system of equations with two tuning parameters can often improve the efficiencies of the estimators beyond those attained by the MDPDEs. If some of these estimators turn out also to be better than the MDPDEs under data contamination (which, as we will see, is indeed true in many cases), then the construction of the generalized  system in Equation (\ref{EQ:new}) will be justified.

To investigate the robustness property of the GBEDE we simulated data from the mixture $(1 - \epsilon) N(0, 1) + \epsilon N(3, 1)$. Here the second component is considered to be a contaminating component, and the parameters of the larger component are the targets. The data are still modeled by a $N(\mu, \sigma^2)$ distribution.  Similarly, for the Poisson model the contaminated data are generated from the mixture of $(1-\epsilon) P(2)+\epsilon P(10)$,  where the parameter of the larger component is our target, and the data are still modeled by a $P(\lambda)$ distribution. Our goal is to compare the efficiency of the GBEDEs with the MLE and MDPDEs.  Tables \ref{normal_mu}, \ref{normal_sigma} and \ref{Poisson} in the Supplementary Material present the relative efficiencies of the estimators with respect to the MLE, where the contamination  proportion is taken as 5\%.  These tables show that the efficiency of the MLE is very low compared to the efficiency of  the GBEDEs with large values of $\beta$. In fact, for large values of $\beta$, the MSEs of the GBEDEs hardly show any change under contamination. On the other hand, for these three tables, the MSEs (times $n$) of the MLEs are 3.70, 9.25 and 21.18, respectively, and they are very high compared to the pure data cases.  We further notice that the efficiency of the GBEDE is improved over the MDPDE if we set $\alpha$ at some appropriate negative value for all $\beta \geq 0.5$. These simulations thus demonstrate that for large positive $\beta$ there are many different GBEDEs which better the corresponding MDPDE under contamination as well as under pure data.

We have conducted another set of simulation study by taking a contamination proportion of 10\%. The results, given in Tables \ref{mu10} and \ref{Poisson10} in the supplementary material, show that the MLEs completely break down in this situation. However, the GBEDEs with large values of $\beta$ give  stable MSEs. We  notice  that in GBEDE the tuning parameter $\beta$ controls the robustness of the estimator -- the larger the value of $\beta$ the higher robustness of the estimator. On the other hand, the other tuning parameter $\alpha$ mainly controls the efficiency of the estimator for a given $\beta$, which is preferably negative for the large values of $\beta$.  Empirically, it may be observed that the optimal value of the tuning parameter combination $(\alpha, \beta)$ in terms of the best efficiency versus robustness compromise may be the rectangle $R = \{ (\alpha, \beta) : \alpha \in [-1,-3],\beta \in [0.2, 0.7]\}$.  

On the whole, Tables \ref{normal_mu_pure}--\ref{Poisson10} in the supplementary material show that there could be several combinations of $(\alpha, \beta)$ values, particularly when $\beta$ is large, which generates GBEDEs that beat the corresponding MDPDEs in terms of both efficiency and outlier stability indicating the benefits of our proposed two parameter system over the DPD and the corresponding estimators. In addition to this we have performed a simulation study to illustrate the suggested root selection strategy and the results are reported at the end of the Supplementary Material.  
  

\begin{figure}
	\centering%
	\includegraphics[height=7cm, width=12cm]{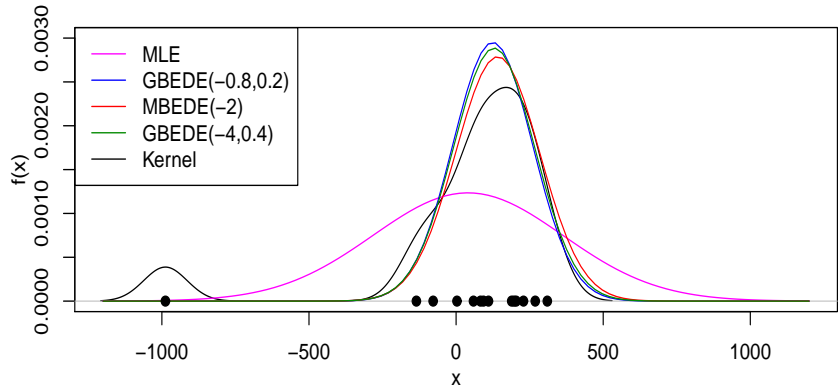}
	\caption{Kernel density estimate and different normal fits for the telephone-Fault data. }
	\label{fig:normal1}
\end{figure}

\section{Real Data Examples} \label{sec:real_data}
\subsection{Telephone-Fault Data}
We consider the data on telephone line faults presented and analyzed by
 \cite{MR909365} and  \cite{MR999667}. The data,
given in Table \ref{TAB:telephone-line-faults}, consist of the ordered
differences between the inverse test rates and the inverse control rates in
14 matched pairs of areas. These data may be modeled as a random sample from a normal distribution with mean $\mu$ and
standard deviation $\sigma$. One fact that is immediately noticeable is that
the first observation of this data set is a huge outlier with respect to the
normal model, while the remaining 13 observations appear to be reasonable
with respect to the same. In Figure \ref{fig:normal1} we present
a kernel density estimate for these data, the normal model fit based on the MLEs of $\mu$ and $\sigma$, and a normal model fit based on the GBEDEs with tuning parameters $\alpha=-2, \ \beta=0.2$.  The MLEs neither model the outlier deleted data, nor provide a fit to the outlier component. On the other hand, the GBEDEs give  excellent fits to the main part of the data. Some GBEDEs for different choice of the tuning parameters are given in Tables  \ref{tele_mu} and \ref{tele_sigma} in the supplementary material. For outlier deleted data, the MLEs of $\mu$ and $\sigma$ are 40.36 and 134.82, respectively, and for the full data, the Huber estimates (with $k=1.5$)  are 103.42 and 149, respectively.  Clearly, all the GBEDEs except those corresponding to $\beta = 0$ are successful in discounting the effect of the large outlier. In fact, there is very little difference in the estimates over $\alpha$ for given fixed values of $\beta$. This shows that choosing a negative value of $\alpha$ has not made the GBEDE any less robust than the corresponding MDPDE in this example.

\begin{table}
	\setlength{\tabcolsep}{4.5pt}
	\caption{Telephone-fault data. }
	\begin{center}
		\setlength{\tabcolsep}{4pt}
		\begin{tabular}{lcccccccccccccc}
			\hline
			Pair & 1 & 2 & 3 & 4 & 5 & 6 & 7 & 8 & 9 & 10 & 11 & 12 & 13 & 14 \\
			Difference & $-988$ & $-135$ & $-78$ & 3 & 59 & 83 & 93 & 110 & 189 & 197 &
			204 & 229 & 269 & 310 \\ \hline
		\end{tabular}
		\label{TAB:telephone-line-faults}
	\end{center}
\end{table}

\subsection{Drosophila Data}
This experiment, considered in \cite{simpson1987minimum}, is known to produce occasional spurious
counts. Male flies were exposed either to a certain degree of chemical to be screened
or to control conditions and the responses are the numbers of recessive lethal mutations among the daughters of such flies. The responses are assumed to be Poisson with mean $\lambda$. The observed and expected frequencies are given in Table \ref{Drosophila_est}. It is seen  that there is an obvious outlier at  91, and as a result  the MLE gives a very poor fit. On the other hand, the GBEDE($-2, 0.4$) and GBEDE($-0.7, 0.1$) or MBEDE($-2$) almost coincide with the outlier deleted MLE (denoted by ``MLE + D''). The GBEDEs corresponding to different values of $\alpha$ and $\beta$ are given in Table \ref{Drosophila_lambda} in the supplementary material. The general observations are similar to the telephone-fault data example.  For comparison we note that the minimum Hellinger distance estimate of  $\lambda$ for these data is 0.36.

\begin{table}
\caption{Different estimators and the corresponding fits for the Drosophila data.}
\begin{center}
\begin{tabular}{l|ccccccc}
\hline
Count &0  & 1 & 2 & 3 & 4 & $\geq 5$ & $\hat{\lambda}$ \\ 
\hline
Observations &  23 & 7 & 3 & 0 & 0 & 1 (91) & -- \\
MLE  & 1.59 & 4.88 & 7.44 & 7.61 & 5.82 & 6.62 & 3.06 \\
MLE + D &  22.93 & 9.03 & 1.78 & 0.23 & 0.02 & 0 & 0.39 \\
GBEDE($-2$,0.4) &  22.79 & 9.12 & 1.82 & 0.24 & 0.02 & 0 & 0.40\\
GBEDE($-0.7$,0.1) &  22.79 & 9.12 & 1.82 & 0.24 & 0.02 & 0 & 0.40\\
 MBEDE($-2$) &  23.72 & 8.54 & 1.54 & 0.18 & 0.01 & 0 & 0.36 \\
\hline
\end{tabular}\label{Drosophila_est}
\end{center} 
\end{table}

\section{Tuning Parameter Selection}\label{sec:tuning}

Selection of the tuning parameter is an important aspect in robust and efficient estimation. Since for the generalized density power divergence (GBEDE), two tuning parameters control the trade off between efficiency and robustness, simultaneous selection is required. \cite{Warjones} proposed a method for determining the tuning parameter of the minimum density power divergence estimator (MDPDE). In this paper we have applied the Warwick and Jones strategy to determine the optimal tuning parameters.


 The  empirical  estimate of the mean square error $MSE(\alpha,\beta)$, as a function of a pilot estimator $\theta^P$,  is given by
\begin{equation}
\widehat{MSE}(\alpha,\beta) = \left( \widehat{\theta}_{\alpha,\beta} - \theta^P\right)^T
\left(\widehat{\theta}_{\alpha,\beta} - \theta^P\right)
+ \frac{1}{n} tr\left(J_{\alpha,\beta}^{-1}(\widehat{\theta}_{\alpha,\beta})
K_{\alpha,\beta}(\widehat{\theta}_{\alpha,\beta})J_{\alpha,\beta}^{-1}(\widehat{\theta}_{\alpha,\beta})\right),
\label{EQ:MSE1}
\end{equation} 
where $\widehat{\theta}_{\alpha,\beta}$ is the GBEDE with tuning parameters $\alpha$ and $\beta$ and $tr(\cdot)$ represents the trace of a matrix. 
 The minimization of this objective function over the tuning parameters  leads to a data driven ``optimal'' estimate  of the tuning parameters.   For the telephone-fault data and drosophila data the optimum tuning parameters of $(\alpha,\beta)$ were found to be  ($-0.8$, 0.2) and ($-0.7$, 0.1), respectively. As suggested by \cite{Warjones} the minimum $L_2$ distance estimator was used as the pilot in the above calculations. In Section \ref{Bregman}, the $L_2$ distance is shown to be a special case of the BED family for $\alpha=0$ in the limiting sense. The minimum $L_2$ distance estimator has strong robustness properties,  which is desired from a pilot estimator.

\section{Application to the Linear Regression} \label{sec:reg}
In this section we consider the extension of the method in the case of linear regression. Our setup consists of  the linear regression model
\begin{equation}
Y_i = x_i^T \gamma + \epsilon_i, \ i = 1,\cdots , n, 
\end{equation}
where $\epsilon_i$'s are independent errors  having $N(0,\sigma^2)$ distributions and  $x_i$'s are fixed design variables. Our data consist of $(Y_i, x_i^T)^T, \ i=1,2, \cdots, n$. Here $\gamma = (\gamma_0, \gamma_1, \cdots, \gamma_p)^T$ is the vector of regression coefficients, and $\theta = (\gamma^T, \sigma^2)^T$ is the set of parameters to be estimated. Note  that $Y_i$'s are independent but not identically distributed as $Y_i \sim f_i(\cdot; \theta)$, where $f_i(\cdot;\theta)$ is  normal with mean $x_i^T \gamma$ and variance $\sigma^2$. In the spirit of Equation (\ref{EQ:new}), we   estimate $\theta$ from the  equation
\begin{equation}
\frac{1}{n}\sum_{i=1}^{n}\left\{ u_i(Y_{i};\theta) f_i^\beta(Y_{i}; \theta) e^{\alpha f_i(Y_{i};\theta)}
-\int_{y_i} u_i(y_i;\theta)f_i^{1+\beta}(y_i;\theta)e^{\alpha f_i(y_i;\theta)}  dy_i \right\}=0, \label{EQ:regression}%
\end{equation}
where $\alpha \in \mathbb{R},\ \beta \geq 0$. 
When $\alpha$ and $\beta$ are equal to zero, the regression estimators corresponding to the above estimating equation are the least square estimators under linear regression with normal errors. \cite{MR3117102} have considered the extension  of the minimum DPD estimator to the  linear regression. As the $(Y_i, x_i^T)^T$ values are not identically distributed, the approach taken therein is to construct the divergence at each individual value given $x_i$, and then average over all the possible values. However, the divergence at each given value of  $x_i$ is based on that single observation only;  see \cite{MR1943184}.
If we put $\alpha=0$ in  estimating equation (\ref{EQ:regression}), we  recover the minimum DPD estimating equation of \cite{MR3117102} in the normal linear regression case. The choice $\alpha =0$ and $\beta =1$ generates the estimating equation of the estimators considered by \cite{MR1943184}.

\subsection{Asymptotic Distribution}
In this section we  derive the asymptotic distribution of  GBEDE $\hat{\theta} = (\hat{\gamma}^T, \hat{\sigma}^2)^T$. We assume that the true data generating distribution belongs to the model family. For the normal error the score function is given by
\begin{equation}
 u_i(y;\theta)  = 
\left( \begin{array}{c}
\frac{(y - x_i^T \gamma)}{\sigma^2} x_i\\
\frac{(y - x_i^T \gamma)^2}{2\sigma^4} - \frac{1}{2\sigma^2}
\end{array} \right).
\end{equation}
Let us define $J_n = \frac{1}{n} \sum_i^n J_i$, and $K_n = \frac{1}{n} \sum_i^n K_i$, where 
\begin{equation}
\begin{split}\label{model_variance2}
& J_i = \int_y u_i(y;\theta)
u_i^T(y;\theta) f_i^{1+\beta}(y;\theta) e^{\alpha f_i(y;\theta)} dy,\\
& K_i = \int_y  u_i(y;\theta)
u_i^T(y;\theta) f_i^{1+2\beta}(y;\theta) e^{2 \alpha f_i(y;\theta)}  dy
- \xi \xi^T, \\
& \xi_i = \int_y u_\theta(x) f_i^{1+\beta}(y;\theta) e^{\alpha f_i(y;\theta)}  dy.
\end{split}
\end{equation}
Under a set of assumptions similar to (A1)--(A7) of \cite{MR3117102}, it can be shown that $\hat{\theta}$ is a consistent estimator of $\theta$. Further, the asymptotic distribution of $\sqrt{n} K_n^{-\frac{1}{2}} J_n (\hat{\theta} -\theta)$ is a multivariate normal with vector mean 0 and covariance matrix $I_p$, the $p$-dimensional identity matrix.


\begin{figure}
\centering
 \includegraphics[height=8cm,width=12cm]{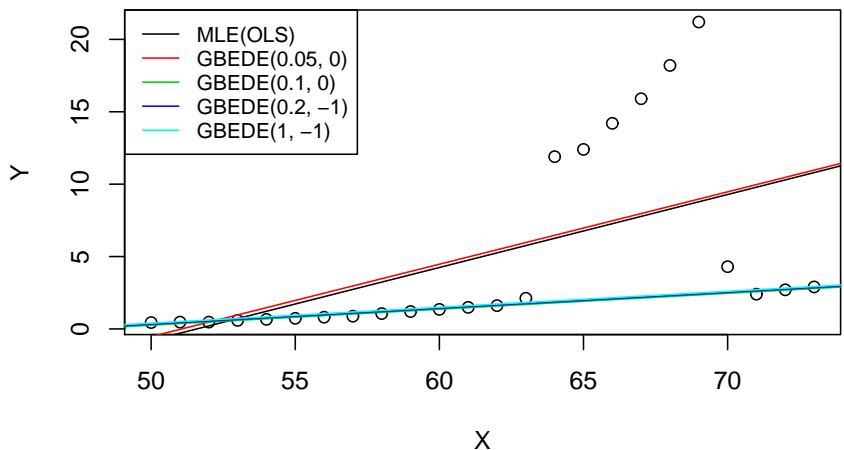}
\caption{Different estimated regression lines for the Belgium telephone call data. }
\label{fig:tele}
\end{figure}

\subsection{Numerical Illustrations}
\subsubsection{Belgium Telephone Call Data}
We consider a real data set from the Belgian Statistical Survey published by the Ministry of Economy of Belgium. The data set is available in \cite{MR914792}, which contains the total number (in tens of millions) of international phone calls made in a year from 1950 to 1973. The scatter plot in Figure \ref{fig:tele} shows a slow upward trend over the years. However, the data include heavy contamination for the period 1964 to 1969 as a result of different recording system being used. The years 1963 and 1970 also display partial effects. On the whole, there are several outliers in the $y$-direction.

\begin{table}
\caption{The GBEDEs and other estimates for the Belgium telephone call data.}
\begin{center}
\begin{tabular}{c|crrrrrrRR}
\hline
$\alpha$ & $\beta$  & 0 & 0.05 & 0.1 & 0.2  & 0.5 & 1 & {\rm Huber} & {\rm Tukey}\\
\hline
\multirow{3}{*}{0}&
$\gamma_0$ & $-26.01$ & $-25.53$ & $-5.24$ & $-5.22$ & $-5.26$ & $-5.36$ & -10.26 & -5.23\\
 & $\gamma_1$ & 0.51 & 0.50 & 0.11 & 0.11 & 0.11 & 0.11 &  0.20 & 0.11\\
 & $\sigma^2$ & 31.61 & 29.17 & 0.02 & 0.01 & 0.01 & 0.02 & 0.82 & 0.03\\
\hline
$\alpha$ & $\beta$  & 0 & 0.05 & 0.1 & 0.2  & 0.5 & 1 & {\rm MM} & {\rm MLE+D}\\
\hline
\multirow{3}{*}{$-1$} &
$\gamma_0$ & $-26.09$ & $-25.66$ & $-5.23$ & $-5.17$ & $-5.08$ & $-5.10$ & -5.24 & -6.35\\
& $\gamma_1$ & 0.51 & 0.50 & 0.11 & 0.11 & 0.11 & 0.11 & 0.11 & 0.13\\
& $\sigma^2$ & 29.00 & 29.22 & 0.02 & 0.02 & 0.01 & 0.01 & 0.04 & 0.04\\
\hline
\end{tabular}
\end{center} 
\label{alpha0}
\end{table}

The ordinary least squares (OLS) estimators of the regression parameters,  lead to the fitted regression line $\hat{Y} = - 26.01 + 0.51 x$ (the $\alpha = 0$, $\beta = 0$, combination in Table \ref{alpha0}). This fitted regression line is plotted in Figure \ref{fig:tele}. It is clear that the MLE is very severely affected by the $Y$-values associated with the years 1964--1969. As a result, the regression line has a large slope and neither fits the main part of the data nor models the outlying data points. 

The GBEDEs of the model parameters for several other ($\alpha, \beta$) combinations  are presented in Table \ref{alpha0}; some of the fitted lines are also plotted in Figure \ref{fig:tele}. The GBEDEs for $\beta \geq 0.1$ exhibit strong robustness properties against  outliers, and give excellent fits to the main structure of the data. Table \ref{alpha0} shows that these estimators are close to the corresponding least squares estimators for the outlier deleted data (denoted by MLE+D), or MM estimators of the parameters for the full data. Table \ref{alpha0} also reports Huber and Tukey's bisquare estimates. We used default tuning parameters for Huber ($k = 1.345$), Tukey's bisquare ($c = 4.685$) and MM estimators as given in `rlm' function of `MASS' package in R. For the MM estimator, the tuning parameter is set to $k_0 = 1.548$; this gives (for $n \gg p$) a breakdown point of 0.5. Table \ref{supply:alpha0} in the Supplementary Material provides the Huber and Tukey's bisquare estimates for different values of the tuning parameters.  

\begin{figure}
	\begin{center}
		\begin{tabular}{rr}
			\includegraphics[height=6cm, width=6cm]{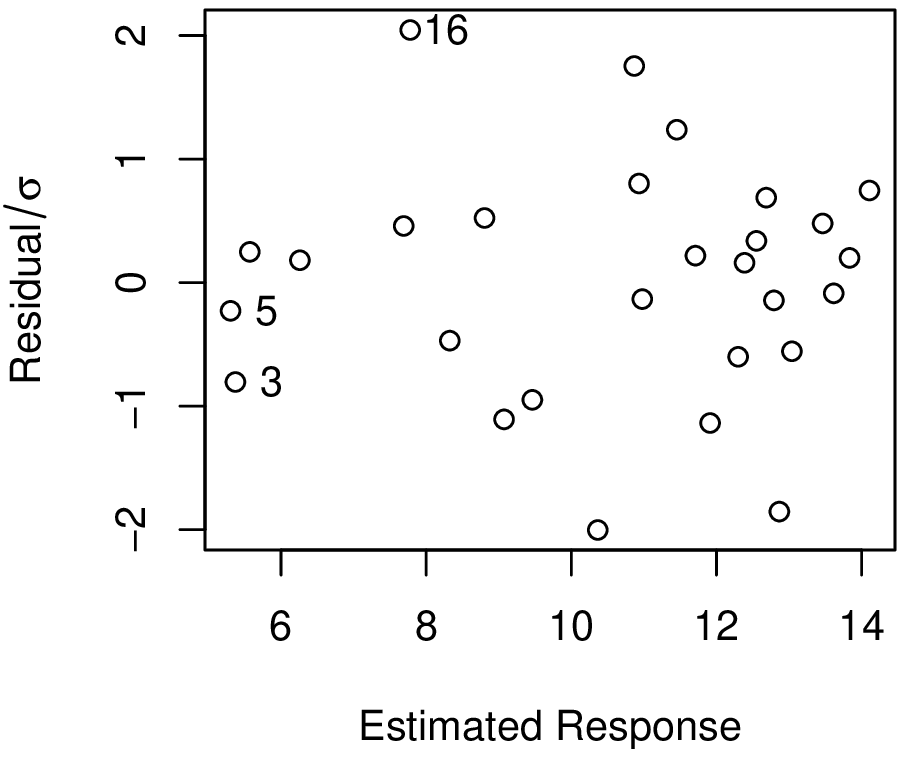}
			&
			\includegraphics[height=6cm, width=6cm]{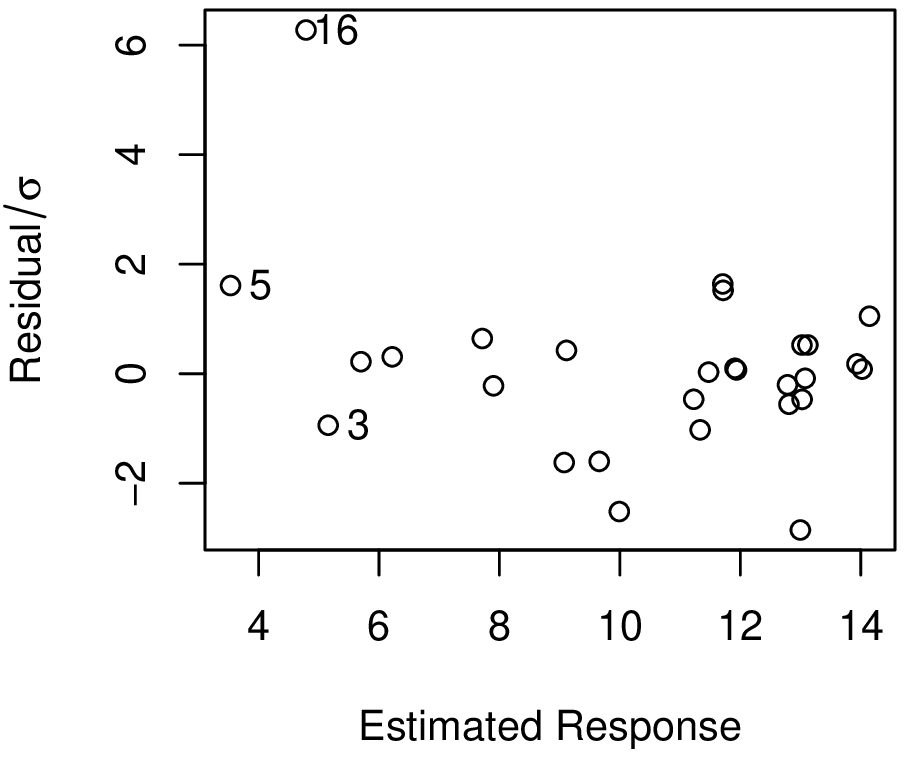}
			\\
			\multicolumn{1}{c}{\textbf{(a)}} & \multicolumn{1}{c}{\textbf{(b)}} \\
		\end{tabular}
		\caption{Residual plot for Salinity data using (a) OLS estimator and (b) GBEDE($-1, 0.5$). }
		\label{fig:salinity}
	\end{center}
\end{figure}

\subsection{Salinity Data}
We now look at a multiple regression example  based on ``Salinity data" given in \cite{MR914792}. The data set contains measurements of water salinity  and river discharge taken in North Carolina's Pamlico Sound, measured  biweekly during the spring seasons of 1972 through 1977. Salinity ($y$) is modeled by salinity lagged by two weeks ($x_1$),  the number of biweekly periods elapsed since the beginning of  spring  ($x_2$), and the volume of river discharge into the sound ($x_3$). The detailed background as well as the statistical challenges in analyzing these  data are described in \cite{carroll1985transformations}.  They indicated that observations 5 and 16 correspond to periods of very heavy discharge. However,  OLS regression fails to detect them as the standardized residuals for all observations are within the accepted range; see Figure \ref{fig:salinity}(a).  On the other hand, Figure \ref{fig:salinity}(b) clearly shows that GBEDEs with $\alpha=-1$ and $\beta=0.5$ detect observation 16 as an influential point, where  the standardized residual is above 6.  \cite{carroll1985transformations} showed that observations 3 and 16
mask the effect of observation 5, so it is not detected at the first stage of analysis. In fact, observation 5 can be recognized as influential only after the deletion of observations 3 and 16.  Table \ref{alpha_beta} reports the values of different GBEDEs and other robust estimators as mentioned in the previous example (additional results are provided in Table \ref{suppli:alpha_beta} of the Supplementary Material). In MLE + D, we deleted observations 5 and 16. As $\beta$ increases, we observe increasing robustness in  the GBEDEs.

\begin{table}
\caption{The GBEDEs and other estimates for Salinity data.}
\begin{center}
\begin{tabular}{c|cRRRRRRRR}
\hline
$\alpha$ & $\beta$  & 0 & 0.05 & 0.1 & 0.2  & 0.5 & 1 & {\rm Huber} & {\rm Tukey}\\
\hline
\multirow{3}{*}{0}
 & $\gamma_0$ & 9.59 & 9.96 & 10.51 & 16.73 & 18.40 & 19.19 & 13.37 & 14.63\\
 & $\gamma_1$ & 0.78 & 0.78 & 0.77 & 0.72 & 0.72 & 0.71 & 0.76 & 0.72\\
 & $\gamma_2$ & -0.03 & -0.03 & -0.04 & -0.15 & -0.20 & -0.18 & -0.09 & -0.10\\
 & $\gamma_3$ & -0.30 & -0.31 & -0.33 & -0.56 & -0.63 & -0.66 & -0.44 & -0.47\\
 & $\sigma^2$ & 1.52 & 1.51 & 1.49 & 1.05 & 0.75 & 0.50 & 0.69 & 0.44\\
\hline
$\alpha$ & $\beta$  & 0 & 0.05 & 0.1 & 0.2  & 0.5 & 1 & {\rm MM} & {\rm MLE+D}\\
\hline
\multirow{3}{*}{$-1$} &
$\gamma_0$ & 9.07 & 9.30 & 9.62 & 11.19 & 18.23 & 18.00 & 18.39 & 23.39\\
& $\gamma_1$ & 
0.78 & 0.78 & 0.78 & 0.77 & 0.72 & 0.72 & 0.71 & 0.70\\
& $\gamma_2$ & 
-0.02 & -0.02 & -0.03 & -0.06 & -0.20 & -0.16 & -0.18 & -0.25\\
& $\gamma_3$ & 
-0.27 & -0.28 & -0.30 & -0.36 & -0.62 & -0.61 & -0.63 & -0.84\\
& $\sigma^2$ & 
1.54 & 1.54 & 1.54 & 1.47 & 0.83 & 0.57 & 1.00 & 0.69\\
\hline
\end{tabular}
\end{center} 
\label{alpha_beta}
\end{table}

For the case of independent, non-homogeneous data, which cover the normal regression case,  \cite{GHOSH} have extended the idea of \cite{Warjones} and have obtained optimal tuning parameter for estimation of regression parameter by minimum density power divergence estimator. The same idea can be applied for the GBEDE for the linear regression model. The extension is relatively straightforward so we do not repeat it here.

\section{Concluding Remarks} \label{sec:conclusion}
In this paper, we have presented a new class of estimators based on Br{\`e}gman class of divergence, which can be expanded to  larger, generalized  set of estimators that contains the MDPDEs  as a special case. 
The striking  observation is that the generalized family can produce, as demonstrated in our numerical studies,  better compromises between robustness and efficiency than what can be provided by the MDPDE alone. A data driven selection of the optimal tuning parameter is also considered.  The idea of this estimator is extended to a non-homogeneous setup, and this estimation procedure is applied to the linear regression model. On the whole the new family of estimators contain several members which can provide better inference than the MDPDEs. 


\bibliographystyle{abbrvnat}
\bibliography{reference}

\newpage
\setcounter{page}{1} 

\begin{center}

{\bf \Large Supplementary Material for ``The B-exponential Divergence and its Generalizations with Applications to Parametric Estimation"}

\end{center}
\setcounter{figure}{0} 
\setcounter{table}{0} 

\section*{Section S1. Generalized Basu et al. Conditions}
\begin{itemize}
 \item[(D1)] The model distributions $F_{\theta }$ of $X$ have common support, so that the set $\mathcal{X} =
\{x|f_{\theta }(x) > 0\}$ is independent of $\theta$. The true distribution $G$ is also supported
on $\mathcal{X}$, on which the corresponding density $g$ is greater than zero.

 \item[(D2)] There is an open subset of $\omega$ of the parameter space $\Theta$, containing the best fitting parameter $\theta _0$ such that for almost all $x \in \mathcal{X}$, and all $\theta  \in \omega$,
the density $f_{\boldsymbol{\theta }} (x)$ is three times differentiable with respect to $\theta $ and the
third partial derivatives are continuous with respect to $\theta$.
 \item[(D3)] The integrals $\int \exp(\alpha f_\theta(x)) f_\theta^{1+\beta}(x) dx$ and$ \int \exp(\alpha f_\theta(x)) f_\theta^{\beta}(x) g(x) dx$ can be differentiated three
times with respect to $\theta$, and the derivatives can be taken under the
integral sign.

 \item[(D4)] 
 The $p \times p$ matrix $\boldsymbol{J}_{\alpha,\beta}(\theta)$, defined  in (\ref{EQ:J-matrix1}),
is positive definite. 

 \item[(D5)] There exists a function $M_{jkl} (x)$ such that
$|\nabla_{kl} V_{\theta,j } (x)| \leq M_{jkl} (x)$ for all $\theta \in \omega$,
where $E_g [M_{jkl} (X)] = m_{jkl} < \infty$ for all $j$, $k$ and $l$, where $V_{\theta, j}$ is the $j$-th component of  
\begin{equation}
V_\theta(x) = u_\theta(x) f_{\theta}^\beta(x) e^{\alpha f{\theta }(x)}
-\int_y u_\theta(y)f_{\theta }^{1+\beta}(y)e^{\alpha f_{\theta }(y)}  dy. 
\end{equation}
\end{itemize}

\section*{Section S2. Figure with Weights under the Normal Model}
\begin{figure}[h]
\centering
{\includegraphics[height=5cm, width=12cm]{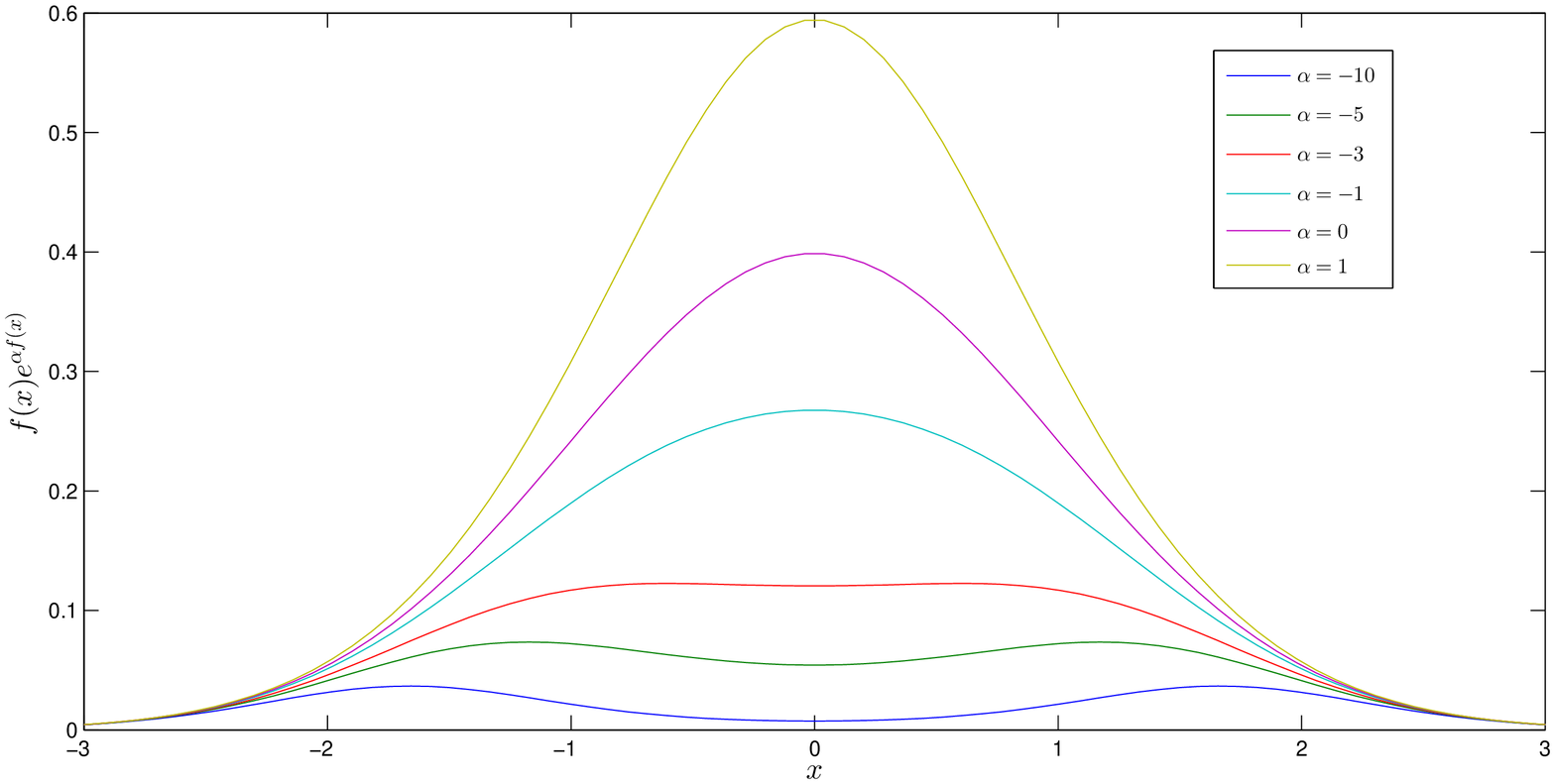}}
\caption{Weights of the score function for the normal model.}
\label{fig:weights}
\end{figure}

\section*{Section S3. Tables for Simulation Results}
\begin{table}[h]
\caption{Efficiency of GBEDEs of $\mu$ under pure normal data.}
\begin{center}
\begin{tabular}{r|ccccccccccc}
\hline
$\alpha \backslash \beta$ &0&0.1  & 0.2&0.2 & 0.4&0.5  & 0.6&0.7&0.8 & 0.9 & 1\\ 
\hline
4 & 0.94 & 0.77 & 0.68 & 0.69 & 0.62 & 0.60 & 0.57 & 0.51 & 0.47 & 0.49 & 0.42\\
 2 & 0.95 & 0.90 & 0.86 & 0.81 & 0.78 & 0.72 & 0.68 & 0.64 & 0.60 & 0.57 & 0.54\\
 0 &  1.00 & 0.99 & 0.96 & 0.92 & 0.89 & 0.85 & 0.80 & 0.76 & 0.72 & 0.69 & 0.65\\
$-1$ & 0.99 & \textbf{1.00} & \textbf{0.99} & \textbf{0.95} & \textbf{0.94} & \textbf{0.89} & \textbf{0.86} & \textbf{0.81} & \textbf{0.77} & \textbf{0.74} & \textbf{0.70}\\
$-2$ & 0.96 & \textbf{0.99} & \textbf{1.00} & \textbf{0.98} & \textbf{0.96} & \textbf{0.94} & \textbf{0.89} & \textbf{0.86} & \textbf{0.81} & \textbf{0.79} & \textbf{0.75}\\
$-3$ & 0.93 & 0.97 & \textbf{0.99} & \textbf{0.98} & \textbf{0.97} & \textbf{0.95} & \textbf{0.93} & \textbf{0.89} & \textbf{0.86} & \textbf{0.83} & \textbf{0.80}\\
$-4$ & 0.89 & 0.94 & \textbf{0.97} & \textbf{0.97} & \textbf{0.97} & \textbf{0.97} & \textbf{0.94} & \textbf{0.92} & \textbf{0.89} & \textbf{0.86} & \textbf{0.83} \\
$-5$ & 0.83 & 0.90 & 0.94 & \textbf{0.95} & \textbf{0.95} & \textbf{0.96} & \textbf{0.94} & \textbf{0.93} & \textbf{0.90} & \textbf{0.88} & \textbf{0.86}\\
$-6$ & 0.79 & 0.86 & 0.90 & \textbf{0.92} & \textbf{0.94} & \textbf{0.95} & \textbf{0.94} & \textbf{0.93} & \textbf{0.91} & \textbf{0.89} & \textbf{0.87}\\
$-7$ & 0.74 & 0.81 & 0.86 & 0.89 & \textbf{0.90} & \textbf{0.94} & \textbf{0.92} & \textbf{0.92} & \textbf{0.91} & \textbf{0.89} & \textbf{0.88}\\
$-8$ & 0.70 & 0.77 & 0.82 & 0.85 & 0.87 & \textbf{0.91} & \textbf{0.90} & \textbf{0.91} & \textbf{0.90} & \textbf{0.89} & \textbf{0.88}\\
\hline
\end{tabular}\label{normal_mu_pure}
\end{center} 
\end{table}

\begin{table}[h]
\caption{Efficiency of GBEDEs of $\sigma$ under pure normal data.}
\begin{center}
\begin{tabular}{r|ccccccccccc}
\hline
$\alpha \backslash \beta$ &0&0.1  & 0.2&0.2 & 0.4&0.5  & 0.6&0.7&0.8 & 0.9 & 1\\ 
\hline
 4 & 0.80 & 0.72 & 0.64 & 0.59 & 0.56 & 0.51 & 0.49 & 0.48 & 0.46 & 0.44 & 0.43\\
 2 & 0.94 & 0.88 & 0.80 & 0.71 & 0.67 & 0.61 & 0.58 & 0.54 & 0.53 & 0.49 & 0.48\\
 0 & 1.00 & 0.98 & 0.91 & 0.84 & 0.77 & 0.72 & 0.66 & 0.62 & 0.58 & 0.56 & 0.53 \\
$-1$ & \textbf{1.00} & \textbf{0.98} & \textbf{0.94} & \textbf{0.88} & \textbf{0.82} & \textbf{0.77} & \textbf{0.71} & \textbf{0.66} & \textbf{0.62} & \textbf{0.59} & \textbf{0.56}\\
$-2$ & 0.98 & \textbf{0.98} & \textbf{0.94} & \textbf{0.91} & \textbf{0.84} & \textbf{0.80} & \textbf{0.74} & \textbf{0.70} & \textbf{0.66} & \textbf{0.63} & \textbf{0.59} \\
$-3$ & 0.94 & 0.96 & \textbf{0.94} & \textbf{0.91} & \textbf{0.88} & \textbf{0.83} & \textbf{0.78} & \textbf{0.74} & \textbf{0.70} & \textbf{0.65} & \textbf{0.62}\\

$-4$ & 0.92 & 0.94 & \textbf{0.94} & \textbf{0.91} & \textbf{0.89} & \textbf{0.86} & \textbf{0.79} & \textbf{0.77} & \textbf{0.72} & \textbf{0.68} & \textbf{0.64}\\
$-5$ & 0.89 & 0.92 & \textbf{0.92} & \textbf{0.91} & \textbf{0.89} & \textbf{0.86} & \textbf{0.80} & \textbf{0.78} & \textbf{0.74} & \textbf{0.70} & \textbf{0.66} \\
$-6$ & 0.88 & 0.91 & \textbf{0.91} & \textbf{0.89} & \textbf{0.89} & \textbf{0.86} & \textbf{0.82} & \textbf{0.79} & \textbf{0.75} & \textbf{0.71} & \textbf{0.68}\\
$-7$ & 0.84 & 0.89 & 0.89 & \textbf{0.89} & \textbf{0.88} & \textbf{0.86} & \textbf{0.82} & \textbf{0.79} & \textbf{0.77} & \textbf{0.72} & \textbf{0.69} \\
$-8$ & 0.83 & 0.86 & 0.88 & \textbf{0.88} & \textbf{0.88} & \textbf{0.86} & \textbf{0.82} & \textbf{0.79} & \textbf{0.77} & \textbf{0.73} & \textbf{0.70}\\
\hline
\end{tabular}\label{normal_sig_pure}
\end{center} 
\end{table}

\begin{table}[h]
\caption{Efficiency of GBEDEs of $\lambda$ under pure Poisson data.}
\begin{center}
\begin{tabular}{r|ccccccccccc}
\hline
$\alpha \backslash \beta$ &0&0.1  & 0.2&0.2 & 0.4&0.5  & 0.6&0.7&0.8 & 0.9 & 1\\ 
\hline
4 & 0.89 & 0.87 & 0.81 & 0.76 & 0.74 & 0.71 & 0.64 & 0.63 & 0.59 & 0.56 & 0.49 \\
 2 &  0.96 & 0.95 & 0.88 & 0.85 & 0.82 & 0.80 & 0.73 & 0.72 & 0.69 & 0.66 & 0.62\\
 0 & 1.00 & 0.99 & 0.94 & 0.90 & 0.88 & 0.86 & 0.79 & 0.79 & 0.76 & 0.73 & 0.70 \\
$-1$ & 0.97 & \textbf{1.00} & \textbf{0.95} & \textbf{0.92} & \textbf{0.91} & \textbf{0.89} & \textbf{0.82} & \textbf{0.82} & \textbf{0.79} & \textbf{0.76} & \textbf{0.73} \\
$-2$ & 0.96 & \textbf{1.00} & \textbf{0.96} & \textbf{0.94} & \textbf{0.92} & \textbf{0.91} & \textbf{0.84} & \textbf{0.85} & \textbf{0.81} & \textbf{0.79} & \textbf{0.76} \\
$-3$ & 0.93 & \textbf{0.99} & \textbf{0.96} & \textbf{0.94} & \textbf{0.94} & \textbf{0.92} & \textbf{0.86} & \textbf{0.87} & \textbf{0.83} & \textbf{0.81} & \textbf{0.79} \\
$-4$ & 0.90 & 0.98 & \textbf{0.95} & \textbf{0.94} & \textbf{0.94} & \textbf{0.93} & \textbf{0.87} & \textbf{0.89} & \textbf{0.85} & \textbf{0.83} & \textbf{0.81}\\
$-5$ & 0.87 & 0.95 & \textbf{0.94} & \textbf{0.94} & \textbf{0.94} & \textbf{0.94} & \textbf{0.88} & \textbf{0.90} & \textbf{0.86} & \textbf{0.85} & \textbf{0.83}\\
$-6$ & 0.84 & 0.92 & 0.92 & \textbf{0.92} & \textbf{0.94} & \textbf{0.94} & \textbf{0.88} & \textbf{0.91} & \textbf{0.87} & \textbf{0.86} & \textbf{0.85} \\
$-7$ & 0.80 & 0.90 & 0.90 & \textbf{0.91} & \textbf{0.93} & \textbf{0.93} & \textbf{0.88} & \textbf{0.91} & \textbf{0.88} & \textbf{0.87} & \textbf{0.86}\\
$-8$ & 0.76 & 0.86 & 0.88 & 0.89 & \textbf{0.92} & \textbf{0.92} & \textbf{0.88} & \textbf{0.91} & \textbf{0.88} & \textbf{0.87} & \textbf{0.86} \\
\hline
\end{tabular}\label{poi_pure}
\end{center} 
\end{table}

\begin{table}[h]	
\caption{Efficiency of GBEDEs of $\mu$ under $5\%$ contaminated normal data.}
\begin{center}
\begin{tabular}{r|ccccccccccc}
\hline
$\alpha \backslash \beta$ &0&0.1  & 0.2 &0.3 & 0.4&0.5  & 0.6&0.7&0.8 & 0.9 & 1\\ 
\hline
4 & \textbf{1.79} & \textbf{2.16} & \textbf{2.24} & 2.00 & 2.09 & 1.89 & 1.81 & 1.86 & 1.67 & 1.41& 1.38\\
2 & \textbf{1.37} & \textbf{1.89} & \textbf{2.14} & \textbf{2.30} & 2.36 & 2.23 & 2.18 & 2.08 & 1.96 & 1.88 & 1.78\\
0 & 1.00 & 1.54 & 1.97 & 2.30 & 2.50 & 2.47 & 2.45 & 2.37 & 2.28 & 2.20 & 2.13\\
$-1$ & 0.85 & 1.35 & 1.80 & 2.20 & 2.47 & \textbf{2.48} & \textbf{2.53} &  \textbf{2.48} & \textbf{2.42} & \textbf{2.34} & \textbf{2.27} \\
$-2$ & 0.73 & 1.17 & 1.62 & 2.07 & 2.37 & 2.43 & \textbf{2.55} & \textbf{2.53} & \textbf{2.50} & \textbf{2.45} & \textbf{2.40}\\
$-3$ & 0.63 & 1.01 & 1.44 & 1.90 & 2.23 & 2.36 & \textbf{2.53} & \textbf{2.55} & \textbf{2.53} & \textbf{2.52} & \textbf{2.50}\\
$-4$ &  0.56 & 0.88 & 1.24 & 1.75 & 1.96 & 2.26 & \textbf{2.64} & \textbf{2.61} & \textbf{2.66} & \textbf{2.50} & \textbf{2.45}\\
$-5$ & 0.54 & 0.88 & 1.27 & 1.71 & 2.07 & 2.24 & \textbf{2.45} & \textbf{2.52} & \textbf{2.55} & \textbf{2.55} & \textbf{2.57}\\
$-6$ & 0.48 & 0.75 & 1.11 & 1.42 & 1.97 & 2.18 & 2.23 & 2.30 & \textbf{2.55} & \textbf{2.72} & \textbf{2.70}\\

$-7$ & 0.38 & 0.60 & 0.88 & 1.23 & 1.55 & 1.78 & 2.06 & 2.23 & \textbf{2.36} & \textbf{2.43} & \textbf{2.53} \\
$-8$ & 0.35 &  0.54 &  0.78 &  1.10 &  1.40 &  1.63 &  1.91 &  2.09 &  2.24 &  \textbf{2.34} & \textbf{2.47}\\
\hline
\end{tabular}\label{normal_mu}
\end{center} 
\end{table}

\begin{table}[h]
\caption{Efficiency of GBEDEs of $\sigma$ under $5\%$ contaminated normal data.}
\begin{center}
\begin{tabular}{r|ccccccccccc}
\hline
$\alpha \backslash \beta$ &0&0.1  & 0.2 &0.3 & 0.4&0.5  & 0.6&0.7&0.8 & 0.9 & 1\\ 
\hline
4 & \textbf{1.96} & \textbf{4.53} & \textbf{6.56} & \textbf{7.23} & 7.06 & 6.95 & 6.80 & 6.75 & 6.51 & 6.42 & 6.25\\
 2 & \textbf{1.38} & \textbf{3.52} & \textbf{5.97} & \textbf{7.58} & \textbf{7.97} & 7.97 & 7.84 & 7.46 & 7.34 & 7.06 & 6.75 \\
 0 & 1.00 & 2.56 & 5.32 & 7.12 & 7.40 & 8.81 & 8.73 & 8.64 & 7.84 & 7.71 & 7.58\\
$-1$  &        0.89 &  2.22 &  4.60 &  \textbf{7.12} &  \textbf{8.41} &  \textbf{8.98} &  \textbf{9.07} &  \textbf{8.81} &  \textbf{8.73} &  \textbf{8.33} &  \textbf{7.91} \\
$-2$ & 0.80 &  1.92 &  4.09 &  6.75 &  \textbf{8.33} &  \textbf{9.16} &  \textbf{9.34} &  \textbf{9.16} &  \textbf{9.16} &  \textbf{8.73} &  \textbf{8.33} \\
$-3$ & 0.72 &  1.68 &  3.64 &  6.29 &  \textbf{8.04} &  \textbf{9.07} &  \textbf{9.44} &  \textbf{9.44} &  \textbf{9.54} &  \textbf{9.16} & \textbf{8.73}\\

$-4$ & 0.65 &  1.48 &  3.25 &  5.82 &  \textbf{7.71} &  \textbf{8.98} &  \textbf{9.44} &  \textbf{9.54} &  \textbf{9.84} &  \textbf{9.44} &  \textbf{9.07} \\
$-5$ & 0.60 &  1.32 &  2.90 &  5.35 & 7.34 &  8.73 &  \textbf{9.44} &  \textbf{9.64} &  \textbf{9.95} &  \textbf{9.64} & \textbf{9.34} \\
$-6$ & 0.56 &  1.19 &  2.59 &  4.89 &  6.90 &  8.41 &  \textbf{9.25} &  \textbf{9.54} &  \textbf{10.05} &  \textbf{9.84} &  \textbf{9.54}\\
$-7$ & 0.52 &  1.08 &  2.34 &  4.49 &  6.51 &  8.11 &  \textbf{8.98} &  \textbf{9.44} &  \textbf{10.05} & \textbf{9.84} &  \textbf{9.64}\\
$-8$ & 0.49 &  0.98 &  2.11 &  4.13 &  6.09 &  7.71 &  \textbf{8.73} &  \textbf{9.25} &  \textbf{9.95} &  \textbf{9.84} &  \textbf{9.74} \\
\hline
\end{tabular}\label{normal_sigma}
\end{center} 
\end{table}

\begin{table}[h]
\caption{Efficiency of GBEDEs of $\lambda$ under $5\%$ contaminated Poisson data.}
\begin{center}
\begin{tabular}{r|ccccccccccc}
\hline
$\alpha \backslash \beta$ &0&0.1  & 0.2 &0.3 & 0.4&0.5  & 0.6&0.7&0.8 & 0.9 & 1\\ 
\hline
4 & \textbf{1.90} &  \textbf{4.64} &  \textbf{6.07} &  6.62 &  6.60 &  6.40 &  6.38 &  5.90 &  5.87 &  5.07 & 4.48\\
 2 & \textbf{1.38} &  \textbf{3.99} &  \textbf{6.05} &  6.99 &  7.20 &  7.08 &  7.08 &  6.66 &  6.64 & 6.03  &  6.00\\
 0 & 1.00 &  3.20 &  5.68 &  7.04 &  7.56 &  7.62 &  7.65 &  7.33 &  7.28 & 6.83 &  6.70\\
$-1$ & 0.86 &  2.82 &  5.36 &  6.94 &  \textbf{7.65} &  \textbf{7.79} &  \textbf{7.84} &  \textbf{7.62} &  \textbf{7.54} &  \textbf{7.16} &  \textbf{7.04}\\
$-2$ & 0.74 &  2.46 &  5.01 &  6.77 &  \textbf{7.65} &  \textbf{7.90} &  \textbf{7.99} &  \textbf{7.84} &  \textbf{7.79} &  \textbf{7.43} & \textbf{7.30}\\
$-3$ & 0.64 &  2.14 &  4.61 &  6.52 &  \textbf{7.59} &  \textbf{7.93} &  \textbf{8.08} &  \textbf{7.99} &  \textbf{7.93} & \textbf{7.67}  & \textbf{7.56} \\
$-4$ & 0.56 & 1.86 &  4.20 &   6.21 &  7.46 &  \textbf{7.90} &  \textbf{8.11} &  \textbf{8.15} & \textbf{8.21} &  \textbf{7.84} &  \textbf{7.76} \\
$-5$ & 0.50 & 1.62 &  3.80 &  5.85 &  7.25 &  \textbf{7.84} &  \textbf{8.11} &  \textbf{8.18} &  \textbf{8.31} &  \textbf{7.99} &  \textbf{7.93} \\
$-6$ & 0.44 & 1.41 &  3.42 &  5.49 &  6.99 &  \textbf{7.70} &  \textbf{8.05} &  \textbf{8.21} &  \textbf{8.34} &  \textbf{8.08} &  \textbf{8.05} \\
$-7$ & 0.39 & 1.23 &  3.06 &  5.10 &  6.70 &  7.54 &  \textbf{7.93} &  \textbf{8.15} &  \textbf{8.34} &  \textbf{8.11} &  \textbf{8.11}\\
$-8$ & 0.36 & 1.08 &  2.73 &  4.71 &  6.40 &  7.30 &  \textbf{7.79} &  \textbf{8.08} &  \textbf{8.31} &  \textbf{8.11} & \textbf{8.18} \\
\hline
\end{tabular}\label{Poisson}
\end{center} 
\end{table}

\begin{table}[h]
\caption{Efficiency of GBEDEs of $\mu$ under $10\%$ contaminated normal data. }
\begin{center}
\begin{tabular}{r|ccccccccccc}
\hline
$\alpha \backslash \beta$ &0&0.1  & 0.2 &0.3 & 0.4&0.5  & 0.6&0.7&0.8 & 0.9 & 1\\ 
\hline

4 & \textbf{2.08} & \textbf{3.40} & \textbf{4.19} & \textbf{4.16} & 4.30 & 4.40 & 4.14 & 4.13 & 4.16 & 4.02 & 3.98 \\
 2 & \textbf{1.48} & \textbf{2.37} & \textbf{3.68} & \textbf{4.19} & \textbf{5.15} & 4.69 & 4.97 & 5.14 & 4.63 & 4.50 & 4.42 \\
 0 &  1.00 & 1.62 & 2.65 & 3.49 & 4.58 & 4.34 & 5.13 & 5.22 & 5.16 & 5.03 & 4.97\\
$-1$ &  0.83 & 1.34 & 2.21 & 3.06 & 4.11 & 4.01 & 5.00 & \textbf{5.38} & \textbf{5.28} & \textbf{5.16} & \textbf{5.15} \\
$-2$ &   0.71 & 1.13 & 1.84 & 2.64 & 3.62 & 3.63 & 4.74 & 5.18 & \textbf{5.27} & \textbf{5.24} & \textbf{5.18}\\
$-3$ &  0.62 & 0.96 & 1.54 & 2.26 & 3.13 & 3.24 & 4.39 & 4.88 & 5.14 & \textbf{5.09} & \textbf{5.25}\\
$-4$ & 0.54 & 0.82 & 1.31 & 1.94 & 2.70 & 2.86 & 3.99 & 4.50 & 4.91 & 4.90 & \textbf{5.16}\\
$-5$ & 0.48 & 0.71 & 1.09 & 1.57 & 2.37 & 2.87 & 3.51 & 3.95 & 4.40 & 4.65 & \textbf{5.16} \\
$-6$ & 0.43 & 0.67 & 0.98 & 1.36 & 1.86 & 2.38 & 3.19 & 3.55 & 4.08 & 4.72 & 4.87 \\
$-7$ &     0.41 & 0.56 & 0.86 & 1.22 & 1.72 & 2.18 & 2.90 & 3.37 & 4.05 & 4.18 & 4.80 \\
$-8$ &   0.37 & 0.52 & 0.78 & 1.04 & 1.49 & 1.84 & 2.28 & 3.02 & 3.31 & 3.96 & 4.83
\\
\hline
\end{tabular}\label{mu10}
\end{center} 
\end{table}

\begin{table}[h]
\caption{Efficiency of GBEDEs of $\sigma$ under $10\%$ contaminated normal data. }
\begin{center}
\begin{tabular}{r|ccccccccccc}
\hline
$\alpha \backslash \beta$ &0&0.1  & 0.2 &0.3 & 0.4&0.5  & 0.6&0.7&0.8 & 0.9 & 1\\ 
\hline
4  &         \textbf{1.72} &  \textbf{3.95} &  \textbf{6.96} &  \textbf{9.01} & 10.04 & 11.49 & 12.09 & 10.71 & 10.79 & 11.11 & 11.04 \\ 
2  & \textbf{1.21} &  \textbf{2.75} & \textbf{5.19} &  \textbf{8.40} &  9.74 & 11.49 & 12.04 & 10.55 & 11.42 & 12.08 & 11.54 \\
 0  &       0.97 &  1.91 & 4.01 &  6.93 & 10.11 & 12.30 & 13.11 &  12.51 & 13.48 & 12.83 & 12.56 \\
$-1$  &        0.86 &  1.62 &  3.66 &  6.17 &  9.03 & 11.06 & 12.08 & \textbf{13.75} & \textbf{13.74} & \textbf{13.64} & \textbf{13.74}\\
$-2$  &        0.78 &  1.47 &  2.83 & 5.37 &  8.91 & 11.33 & 12.80 & \textbf{13.30} & 13.47 & \textbf{13.99} & \textbf{14.25}\\
$-3$ & 0.71 &  1.25 &  2.46 &  4.98 &  7.64 &  9.91 & 12.86 & \textbf{15.04} &  \textbf{13.67} & \textbf{16.08} & \textbf{14.13} \\
$-4$ & .67 &  1.16 &  2.21 &  4.19 &  6.51 &  9.53 & 12.43 & \textbf{14.18} & \textbf{14.61} & \textbf{15.24} & \textbf{14.93}\\
$-5$  &      0.63 &  0.97 &  1.87 &  3.62 &  6.54 &  9.64 & 12.05 & \textbf{12.99} & \textbf{15.08} & \textbf{14.62} & \textbf{15.27} \\
$-6$   &   0.57 &  0.92 & 1.75 &  3.17 &  5.65 &  7.81 & 12.06 & 12.33 & \textbf{16.67} & \textbf{17.60} & \textbf{15.99}\\
$-7$  &    0.54 &  0.80 &  1.62 &  2.81 &  4.95 &  7.29 & 10.51 & 12.017 & \textbf{15.11} & \textbf{16.68} & \textbf{15.36}\\
$-8$  &   0.52 &  0.76 &  1.46 &  2.51 &  4.62 &  6.75 & 10.00 & 12.56 & \textbf{13.81} & \textbf{15.38} & \textbf{16.89}
\\
\hline
\end{tabular}\label{sigma10}
\end{center} 
\end{table}

\begin{table}[h]
\caption{Efficiency of GBEDEs of $\lambda$ under $10\%$ contaminated Poisson data.  }
\begin{center}
\begin{tabular}{r|ccccccccccc}
\hline
$\alpha \backslash \beta$ &0&0.1  & 0.2& 0.3 & 0.4&0.5  & 0.6&0.7&0.8 & 0.9 & 1\\ 
\hline
4 & \textbf{1.72} &  \textbf{3.95} &  6.96 &  9.01 & 10.04 & 11.49 & 12.09 & 10.71 & 10.79 & 11.11 & 11.04 \\
 2 & \textbf{1.35} &  \textbf{4.57} & \textbf{9.83} & \textbf{14.91} & 18.97 & 18.38 & 18.19 & 17.71 & 17.19 & 15.75 & 16.36\\
 0 & 1.00 &  3.32 &  7.98 & 13.88 & 19.16 & 19.67 & 19.87 & 19.38 & 19.22 & 17.90 & 18.79 \\
$-1$ &  0.86 &  2.82 &  7.04 & 13.06 & 18.86 & \textbf{20.00} & \textbf{20.47} & \textbf{20.03} & \textbf{20.09} & \textbf{19.98} & \textbf{19.94}\\
$-2$ &  0.76 &  2.39 &  6.14 & 12.12 & 18.30 & \textbf{20.11} & \textbf{20.88} & \textbf{20.53} & \textbf{20.82} & \textbf{19.82} & \textbf{20.98} \\
$-3$ & 0.67 &  2.04 &  5.32 & 11.08 & 17.53 & \textbf{19.98} & \textbf{21.07} & \textbf{20.87} & \textbf{21.39} & \textbf{20.61} & \textbf{21.89}\\
$-4$ & 0.60 &  1.75 &  4.58 & 10.02 & 16.58 & 19.62 & \textbf{21.06} & \textbf{21.03} & \textbf{21.80} & \textbf{21.26} & \textbf{22.63} 
 \\
$-5$ & 0.54 &  1.51 &  3.94 &  8.97 & 15.50 & 19.05 & \textbf{20.84} & \textbf{21.02} & \textbf{22.04} & \textbf{21.75} & \textbf{23.19} \\
$-6$ &        0.49 &  1.31 &  3.39 &  7.96 & 14.35 & 18.31 & \textbf{20.45} & \textbf{20.85} & \textbf{22.12} & \textbf{22.08} & \textbf{21.55} \\
$-7$  &     0.45 &  1.15 &  2.92 &  7.02 & 13.16 & 17.43 & 19.86 & \textbf{20.52} & \textbf{22.05} & \textbf{22.25} & \textbf{23.74}\\
$-8$ &       0.41 &  1.01 &  2.53 &  6.17 & 11.97 & 16.44 & 19.21 & \textbf{20.05} & \textbf{21.83} & \textbf{22.28} & \textbf{23.77}\\
 
\hline
\end{tabular}\label{Poisson10}
\end{center} 
\end{table}

\section*{Section S4. Tables for Real Data Examples}
Tables \ref{tele_mu}--\ref{suppli:alpha_beta} provide the real data examples.

\begin{table}[h]
\caption{The GBEDEs of $\mu$ for the Telephone-fault data.}
\begin{center}
\begin{tabular}{r|ccccccc}
\hline
$\alpha \backslash \beta$ &0  & 0.2 & 0.4  & 0.6 & 0.8 & 1\\ 
\hline
4  &  40.66  &  124.47  &  129.35  &  133.75  &  137.81  &  141.48\\
2  &  40.51  &  124.42  &  129.30  &  133.70  &  137.77  &  141.44\\
0  &  40.36  &  124.37  &  129.26  &  133.66  &  137.73  &  141.41\\
$-2$  &  40.21  &  124.32  &  129.21  &  133.62  &  137.69  &  141.38\\
$-4$  &  40.06  &  124.27  &  129.16  &  133.58  &  137.65  &  141.34\\
$-6$  &  39.91  &  124.22  &  129.12  &  133.53  &  137.61  &  141.31\\
$-8$  &  39.76  &  124.18  &  129.07  &  133.49  &  137.58  &  141.28\\
\hline
\end{tabular}\label{tele_mu}
\end{center} 
\end{table}

\begin{table}[h]
\caption{The GBEDEs of $\sigma$ for the Telephone-fault data.}
\begin{center}
\begin{tabular}{r|ccccccc}
\hline
$\alpha \backslash \beta$ &0  & 0.2 & 0.4  & 0.6 & 0.8 & 1\\ 
\hline
4  &  311.24  &  135.23  &  138.29  &  140.73  &  142.24  &  143.12\\
2  &  311.36  &  135.20  &  138.27  &  140.71  &  142.22  &  143.11\\
0  &  311.48  &  135.17  &  138.24  &  140.70  &  142.21  &  143.10\\
$-2$  &  311.60  &  135.14  &  138.22  &  140.68  &  142.19  &  143.08\\
$-4$  &  311.72  &  135.12  &  138.20  &  140.66  &  142.18  &  143.07\\
$-6$  &  311.84  &  135.09  &  138.18  &  140.64  &  142.17  &  143.06\\
$-8$  &  311.95  &  135.06  &  138.15  &  140.63  &  142.15  &  143.04\\
\hline
\end{tabular}\label{tele_sigma}
\end{center} 
\end{table}

\begin{table}[h]
\caption{The GBEDEs of $\lambda$ for the Drosophila data.}
\begin{center}
\begin{tabular}{r|ccccccc}
\hline
$\alpha \backslash \beta$ &0  & 0.2 & 0.4  & 0.6 & 0.8 & 1\\ 
\hline
4  & 2.17  &  0.38  &  0.38  &  0.38  &  0.38  &  0.38\\
2  &   2.73  &  0.38  &  0.38  &  0.38  &  0.38  &  0.38\\
0  &   3.06  &  0.39  &  0.38  &  0.37  &  0.37  &  0.37\\
$-2$ &  3.30  &  0.41  &  0.40  &  0.38  &  0.37  &  0.36\\
$-4$ &  3.48  &  0.43  &  0.42  &  0.41  &  0.39  &  0.37\\
$-6$ &  3.64  &  0.44  &  0.44  &  0.44  &  0.42  &  0.40\\
$-8$ &  3.77  &  0.45  &  0.46  &  0.46  &  0.46  &  0.44\\
\hline
\end{tabular}\label{Drosophila_lambda}
\end{center} 
\end{table}

\begin{table}
\caption{The Huber and Tukey's bisquare estimates  for different values of the tuning parameters for the Belgium Telephone call data.}
\begin{center}
\begin{tabular}{c|cRRRRRRRR}
\hline
Huber & $\beta$  & 2 & 1.75 & 1.5 & 1.345 & 1.25 & 1 & 0.75 & 0.5 \\ 
\hline
\multirow{3}{*}{}&
$\gamma_0$ &  -24.81 & -23.04 & -10.71 & -10.26 & -10.04 & -9.65 & -8.48 & -7.21 \\ 
& $\gamma_1$ &   0.48 & 0.45 & 0.21 & 0.20 & 0.20 & 0.19 & 0.17 & 0.15 \\ 
& $\sigma^2$ &     23.86 & 20.41 & 1.06 & 0.82 & 0.71 & 0.58 & 0.37 & 0.12 \\ \hline
Tukey & $\beta$  &   6 & 5.75 & 5.5 & 5 & 4.685 & 4.5 & 4.25 & 4 \\ 
\hline
\multirow{3}{*}{}&
$\gamma_0$ &  -23.56 & -22.69 & -5.24 & -5.23 & -5.23 & -5.23 & -5.22 & -5.22 \\ 
& $\gamma_1$ &   0.46 & 0.44 & 0.11 & 0.11 & 0.11 & 0.11 & 0.11 & 0.11 \\ 
& $\sigma^2$ &   19.91 & 18.15 & 0.03 & 0.03 & 0.03 & 0.03 & 0.03 & 0.03 \\
\hline
\end{tabular}
\end{center} 
\label{supply:alpha0}
\end{table}

\begin{table}
\caption{The Huber and Tukey's bisquare etimates for different values of the tuning parameters for Salinity data.}
\begin{center}
\begin{tabular}{c|cRRRRRRRR}
\hline
Huber & $\beta$  & 2 & 1.75 & 1.5 & 1.345 & 1.25 & 1.00 & 0.75 & 0.5 \\ 
\hline
\multirow{3}{*}{}
 & $\gamma_0$ &   12.01 & 12.52 & 12.98 & 13.37 & 13.56 & 14.09 & 14.54 & 14.63 \\ 
& $\gamma_1$ &   0.76 & 0.76 & 0.76 & 0.76 & 0.75 & 0.74 & 0.73 & 0.72 \\ 
& $\gamma_2$ &   -0.08 & -0.09 & -0.10 & -0.09 & -0.09 & -0.11 & -0.12 & -0.10 \\ 
& $\gamma_3$ &   -0.39 & -0.41 & -0.42 & -0.44 & -0.45 & -0.46 & -0.47 & -0.47 \\ 
& $\sigma^2$ &   0.82 & 0.80 & 0.79 & 0.69 & 0.64 & 0.57 & 0.45 & 0.44 \\ \hline
Tukey & $\beta$ & 6 & 5.75 & 5.5 & 5 & 4.685 & 4.5 & 4.25 & 4 \\ 
\hline
\multirow{3}{*}{}
 & $\gamma_0$ &   18.38 & 18.37 & 18.36 & 18.32 & 18.25 & 18.23 & 18.18 & 18.13 \\ 
& $\gamma_1$ &   0.71 & 0.71 & 0.72 & 0.72 & 0.73 & 0.73 & 0.73 & 0.73 \\ 
& $\gamma_2$ &   -0.18 & -0.18 & -0.19 & -0.20 & -0.21 & -0.21 & -0.21 & -0.21 \\ 
& $\gamma_3$ &   -0.63 & -0.63 & -0.63 & -0.62 & -0.62 & -0.62 & -0.62 & -0.62 \\ 
& $\sigma^2$ &   0.54 & 0.53 & 0.53 & 0.53 & 0.47 & 0.48 & 0.49 & 0.51 \\
\hline
\end{tabular}
\end{center} 
\label{suppli:alpha_beta}
\end{table}

\begin{figure}
\begin{center}
 \includegraphics[height=5cm, width=15cm]{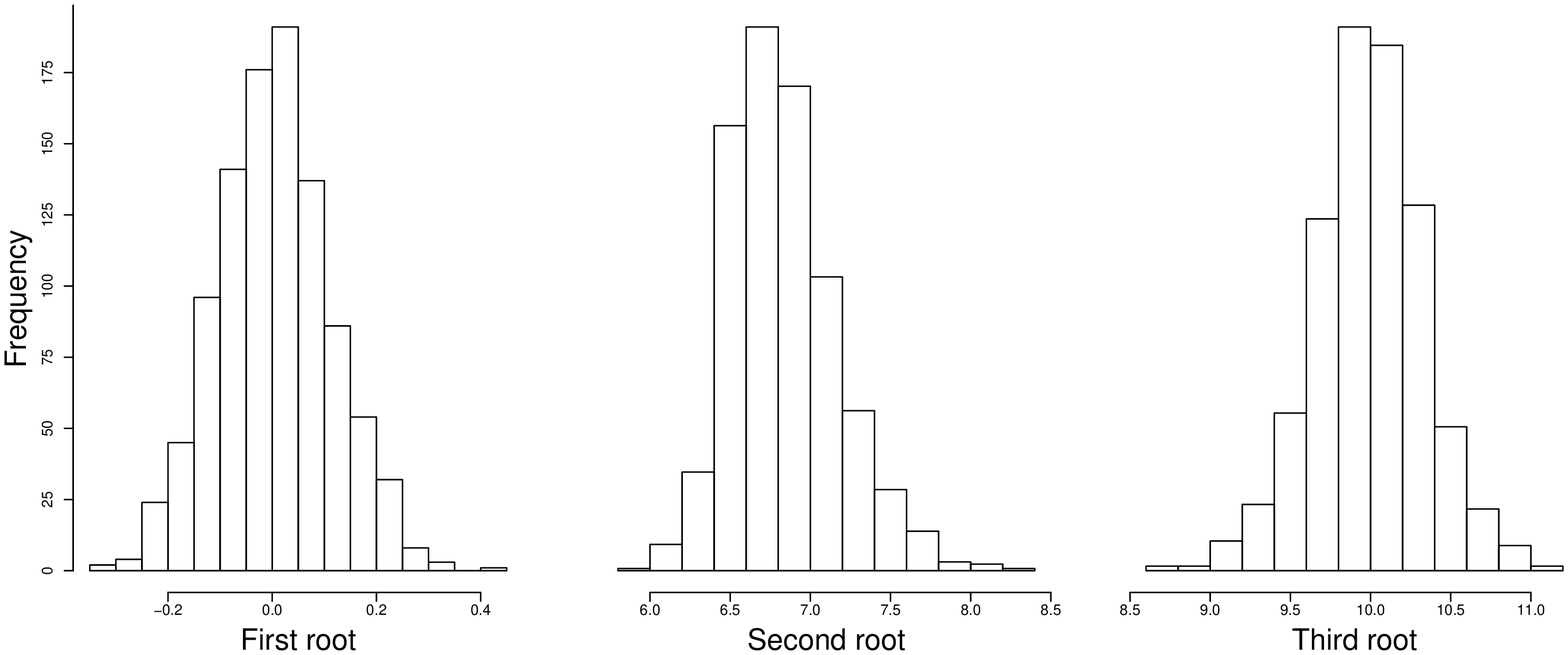}
\caption{The histograms of the three roots of GBEDE$(-1, 0.2)$ in the normal mixture data. }
\label{fig:roots}
\end{center}
\end{figure}

\section*{Section S5. A Root Selection Example} 

In Section 3 of the main paper, we have presented a root selection algorithm in case of multiple roots through the construction of an empirical divergence measure. Here, we illustrate its application on a particular sample of size 30 modeled by a $N(\mu, \sigma^2)$ distribution. The particular sample that we have utilized is presented in Biswas et al. (2015, Table II). In calculating the GBEDEs for these data, we get two distinct roots for the combination $(\alpha, \beta) = (-1, 0.2)$. The roots representing the $(\mu,  \sigma^2)$ parameters are   (0.1122,    0.7935) and  $(-1.4362,  0.0079)$ for this sample. 
The corresponding empirical divergence measures (without the constant term), mentioned in Equation (\ref{div_gen}), are $0.0670$ and $0.1928$, respectively, so that we select the first root. A quick scrutiny shows that the second root essentially concentrates on a small minority of points at the left tail of the data set, and cannot be a reasonable representative of the entire data set.  

We further illustrate the root selection technique using a simulated data. We generated a sample of size 100 from a normal mixture model with known variance.  The target distribution is $N(\mu=0,\sigma=1)$, and 10\% of the data are contaminated with $N(\mu_c=10,\sigma=1)$. We are interested in estimating $\mu$, the mean of the target distribution assuming a single normal distribution with $\sigma=1$. The GBEDE with parameters  $(\alpha, \beta) = (-1, 0.2)$ produces three roots of $\mu$ in each of the 1000 replications performed. Figure \ref{fig:roots} shows the histograms of those three roots over the 1000 replications. In R, `uniroot.all' function of `rootSolve' library is used to solve the estimating equations.  The mean values of the estimates of $\mu$ are 0.0041, 6.8336 and 9.9934, respectively. Note that there are two roots, one each around $\mu_0$ and $\mu_c$, and the third one is somewhere in the middle of them. For these three roots, the average values of the empirical divergence measures, mentioned in Equation (\ref{div_gen}), are $-0.0065$,  0.1577 and  0.1407, respectively. Moreover, in this example, the first root always (i.e. in each of the 1000 replications) produces the lowest value of the divergence measure. Therefore,  this strategy appears to necessarily select the root which represents the target value of the the parameter. 

\section*{Section S6. Sketch of the Proof of Theorem \ref{theorem1}}
Here we provide a sketch of the proof of Theorem \ref{theorem1}. Notice that apart from the score function in the score equation that we are dealing with, we also have a valid objective function as given in Equation (\ref{div_gen}). 
One can then basically mimic Lehmann's (1983) proof for the consistency of the maximum likelihood estimator to establish the consistency of the GBEDE. 
Roughly, this consistency proof proceeds as follows: 

\begin{itemize}

\item Let $G$ be the true distribution, and let $\theta_{\alpha, \beta} = T_{\alpha, \beta}(G)$. Let $H_n(\theta)$ be the objective function in Equation (\ref{div_gen}).  
Consider the behavior of this objective function $H_n(\theta)$ on the surface of a ball of radius $a$, which has $\theta_{\alpha, \beta}$ as its centre. Using a Taylor series expansion of $H_n(\theta)$ around 
$\theta_{\alpha, \beta}$, we control the linear, quadratic and cubic terms appropriately, so that for small enough $a$, $H_n(\theta) - H_n(\theta_{\alpha, \beta})$ is positive for all $\theta$ on the surface of the ball. Hence, there must be at least a local minimizer of the objective function $H_n(\theta)$ within the ball of radius $a$. Since the estimating equation must be satisfied at a local minimum, there must be a root of the estimating equation within the ball. For ease of presentation we will refer to $\theta_{\alpha,\beta}$ as $\theta_0$ is the rest of this discussion.

\item This Taylor series expansion may be expressed as
\begin{eqnarray*}
H_n(\theta_0)- H_n(\theta) & = & - \left\{ \sum_j (\theta_j - \theta_{0j})\nabla_j H_n(\theta)|_{\theta = \theta_0} \right.\\ 
& + & \sum_j \sum_k (\theta_j - \theta_{0j}) (\theta_k - \theta_{0k}) \nabla_{jk} H_n(\theta)|_{\theta = \theta_0}\\
& + & \left. \sum_j \sum_k \sum_l (\theta_j - \theta_{0j}) (\theta_k - \theta_{0k})(\theta_l - \theta_{0l}) \sum_{i=1}^n \gamma_{jkl}(X_i) M_{jkl}(X_i) \right\}\\
& = & S_1 + S_2 + S_3~ (\rm{say}),
\end{eqnarray*}
where $\theta_j$ and $\theta_{0j}$ refer to the indicated components of $\theta$ and $\theta_0$ respectively, $j = 1, 2, \ldots, p$, and 
$0 \leq |\gamma_{jkl}(x)| \leq 1$ by assumption (D5). The proof proceeds by demonstrating that 
\begin{itemize}
\item[(i)] $\nabla_j H_n(\theta)|_{\theta = \theta_0} \underset{n\longrightarrow \infty }{\overset{p}{\longrightarrow }} 0$. 
\item[(ii)] $\nabla_{jk} H_n(\theta)|_{\theta = \theta_0}  \underset{n\longrightarrow \infty }{\overset{p}{\longrightarrow }} -J_{jk}$, where $J_{jk}$
is the $(jk)$-th  element of the matrix $J_{\alpha, \beta}(\theta_0)$, and $J_{\alpha, \beta}(\theta)$ is as defined in Equation (\ref{EQ:J-matrix1}).  
\item[(ii)] $\sum_{i=1}^n \gamma_{jkl}(X_i) M_{jkl}(X_i)$ is bounded, with probability tending to 1. 
\end{itemize}
\end{itemize}
Under the given conditions, the above three results are straightforward. 

The proof of the second part of the theorem, involving the asymptotic  normality of the estimator proceeds exactly in the same way as in the proof of the asymptotic  normality of the maximum likelihood estimator as given by Lehmann (1983, pp. 432--434). 
\section*{Reference}
\begin{itemize}
\item[1] Biswas, A., Roy, T., Majumder, S.,  Basu, A. (2015). A new weighted likelihood approach. Stat, 4(1), 97--107.
\item[2]Lehmann E. L.(1983). Theory of Point Estimation, John Wiley and Sons, New York.

\end{itemize}

\end{document}